\documentclass[fleqn,usenatbib]{mnras}
\UseRawInputEncoding
\DeclareRobustCommand{\VAN}[3]{#2}
\let\VANthebibliography\thebibliography
\def\thebibliography{\DeclareRobustCommand{\VAN}[3]{##3}\VANthebibliography}

\usepackage{graphicx}	% Including figure files
\usepackage{multirow}
\usepackage{subfig}

\newcommand{\hMsun}{ h^{-1}{\rm M_{ \odot}}}
\newcommand{\hMpc}{ h^{-1}{\rm Mpc}}

\newcommand{\ihMpc}{ h\,{\rm Mpc}^{-1}}

\newcommand{\fnu}{ f(\nu)}
\newcommand{\aeff}{ \alpha_{\rm eff}}
\newcommand{\neff}{ n_{\rm eff}}

\title[Non-universality of the mass function]{Non-universality of the mass function: dependence on the growth rate and power spectrum shape}

\author[L. Ondaro-Mallea et al.]{
Lurdes Ondaro-Mallea,$^{1,2}$\thanks{E-mail: lurdes.ondaro@estudiante.uam.es (LO)}
Raul E. Angulo,$^{1,3}$ \thanks{E-mail: reangulo@dipc.org (REA)}
Matteo Zennaro,$^{1}$
Sergio Contreras,$^{1}$
\newauthor
and Giovanni Aric\`o$^{1,4}$.
\\
$^{1}$Donostia International Physics Center (DIPC), Manuel Lardizabal Ibilbidea, 4, 20018 Donostia, Gipuzkoa, Spain\\$^{2}$Universidad Aut\'onoma de Madrid (UAM), C/ Francisco Tom\'as y Valiente, 7, 28049 Madrid, Spain\\
$^{3}$IKERBASQUE, Basque Foundation for Science, 48013, Bilbao, Spain\\
$^{4}$Universidad de Zaragoza, Pedro Cerbuna 12, 50009 Zaragoza, Spain\\ }

\date{Accepted XXX. Received YYY; in original form ZZZ}

\pubyear{2021}

\begin{document}
\label{firstpage}
\pagerange{\pageref{firstpage}--\pageref{lastpage}}
\maketitle

\begin{abstract}
The abundance of dark matter haloes is one of the key probes of the growth of structure and expansion history of the Universe. Theoretical predictions for this quantity usually assume that, when expressed in a certain form, it depends only on the mass variance of the linear density field. However, cosmological simulations have revealed that this assumption breaks, leading to 10-20\% systematic effects. In this paper we employ a specially-designed suite of simulations to further investigate this problem. Specifically, we carry out cosmological $N$-body simulations where we systematically vary growth history at a fixed linear density field, or vary the power spectrum shape at a fixed growth history. We show that the halo mass function generically depends on these quantities, thus showing a clear signal of non-universality. Most of this effect can be traced back to the way in which the same linear fluctuation grows differently into the nonlinear regime depending on details of its assembly history. With these results, we propose a parameterization with explicit dependence on the linear growth rate and power spectrum shape. Using an independent suite of simulations, we show that this fitting function accurately captures the mass function of haloes over cosmologies spanning a vast parameter space, including massive neutrinos and dynamical dark energy. Finally, we employ this tool to improve the accuracy of so-called cosmology-rescaling methods and show they can deliver 2\% accurate predictions for the halo mass function over the whole range of currently viable cosmologies.
\end{abstract}


%%%%%%%%%%%%%%%%%%%%%%%%%%%%%%%%%%%%%%%%%%%%%%%%%%

%%%%%%%%%%%%%%%%% BODY OF PAPER %%%%%%%%%%%%%%%%%%

\section{Introduction}
\label{sec:intro}

Collapsed dark matter structures, a.k.a. haloes, offer an important way to constrain fundamental properties of the Universe. The abundance of haloes is sensitive to the growth of structure and the statistics of primordial fluctuations, thus it can be employed to, for instance, constrain the value of cosmic parameters including dark energy and the sum of neutrino masses \citep{Weinberg2013}.

In the next decades, up to hundreds of thousands of haloes with mass above $\sim 10^{13}\,\hMsun$ will be detected by upcoming observational surveys (e.g. {\it eROSITA} \citep{Hofmann:2017}, {\it EUCLID} \citep{Sartoris:2016}, {\it LSST} \citep{Ivezic:2019}, {\it Simons Observatory} \citep{Ade:2019}, {\it CMB-S4} \citep{Abazajian:2019}, and {\it J-PAS} \citep{Bonoli:2020}). These future surveys will employ various observables over different wavelengths to identify haloes, such as their Sunyaev-Zeldovich effect, X-ray emission, gravitational lensing, or number of optically-detected galaxies. Despite these differences, a necessary ingredient for all such analyses is accurate predictions for the abundance of haloes of a given mass as a function of cosmological parameters.

In the Press-Schechter formalism \citep{PressSchechter:1974} (hereafter, PS), the abundance of dark matter haloes of mass $M$ is fundamentally given by the relative abundance of peaks of different types in a Gaussian random field. Specifically, the halo mass function reads:

\begin{equation}
\label{eq:mass_function}
n(M) d\log(M) = \frac{-1}{3} \frac{\rho_{\rm b}}{M} \frac{d \log \sigma}{d \log M}\,\nu f(\nu)
\end{equation}

\noindent where $\rho_{\rm b}$ is the background matter density of the Universe; $f(\nu) = \sqrt{2/\pi} \exp(-0.5 \nu^2)$; $\nu$ is the so-called ``peak height'' associated to a halo of mass $M$ and is defined as  $\nu \equiv \delta_c(z)/\sigma(M,z)$; $\delta_c$ is the critical overdensity for collapse; and $\sigma(M,z)$ is the rms linear variance extrapolated at the redshift of interest, $z$.

In this approach, cosmological parameters and the shape of the power spectrum of fluctuations, $P(k)$, are considered only through modifications to $\sigma$:
\begin{equation}\label{eq:sigma}
\sigma(R,z) = \frac{D^2(z)}{2\pi}\int_{0}^{\infty} d^3 k P(k) \tilde{W}(k,R)^{2}
\end{equation}

\noindent where $D(z)$ is the linear growth factor, and $\tilde{W}(k,R)$ is the Fourier transform of the top-hat window function and $M=\frac{4\pi}{3}\rho_b R^{3}$. On top of this, $n(M)$ is affected by the cosmologcal parameters through $\rho_b$. Since $f(\nu)$ is cosmology-independent, the halo mass function is predicted to be ``universal''.

The ``universality'' of the mass function is a key property because it allows for accurate predictions even if PS itself is inaccurate. For instance, if the mass function is universal, a {\it single} simulation is needed to measure $f(\nu)$, and then use Eq.~\ref{eq:mass_function} to make predictions for any cosmological model. Therefore, computational resources can be focused on accurately measuring $f(\nu)$ using high force and high mass resolution simulations of large cosmic volumes, rather than requiring large ensembles of simulations spanning the full range of cosmological parameters of interest.

In fact, several early studies found that the PS halo mass function describes only qualitatively the abundance of dark matter haloes in $N$-body simulations. Motivated by the universality of the mass function, these works have provided much more precise fitting functions for $\fnu$, usually employing functional forms inspired by ellipsoidal collapse, but still assuming that all cosmology dependence can be captured through $\sigma(M)$ \citep[e.g.][]{Sheth:1999, Jenkins:2001, Sheth:2002, Reed:2003, Warren:2006, Reed:2007,  Crocce:2010, Bhattacharya:2011, Angulo:2012, Watson:2013, Bocquet:2016, Seppi:2020}.

More recently, various authors pointed out and quantified the ``non-universality'' of the mass function \citep{Tinker:2008,Courtin:2010,Despali:2015,Diemer:2020}. They have found that the amplitude and shape of $\fnu$ does depend on redshift and cosmology in a complicated manner, which depends on the halo definition, and can modify by up to 10\% the expected abundance of haloes of a given mass. This can be easily the leading theory systematic error in the cosmological analysis of future cluster catalogues \citep{artis:2021}.

%Specifically, \cite{Despali:2015} showed that at least some of the non-universality of the mass function is related to the definition of halo mass, and argued that

One of the main goals of this paper is to explore the non-universality of the halo mass function. That is, the dependence of the abundance of dark matter haloes on cosmology and/or redshift in addition to that on the linear rms variance of fluctuations, $\sigma(M)$. For this, we will consider cosmologies with identical values for $\sigma(R)$ at $z=0$, but with very different growth histories. In this way, any signal of non-universality can be attributed to  the way in which haloes grow, since the statistics of the initial fluctuation field are identical. This can shed light on the origin of the mass function non-universality and allow for a more accurate modelling. In addition, we will consider simulations with fixed growth history but varying the power spectrum of primordial fluctuations.

Indeed we will show that by explicitly accounting for the dependence of $f(\nu)$ on the growth rate and power spectrum slope, we are able to predict the halo mass function with a 2-3\% accuracy over essentially the whole currently viable cosmological parameter space, including dynamical dark energy. Moreover, this modelling allows to improve the accuracy with which cosmology-rescaling algorithms predict the abundance of haloes.

The outline of this paper is as follows. In \S\ref{sec:simulations} we describe the cosmological models and the respective $N$-body simulation we carry out. In \S\ref{sec:nonuniversality} we measure the non-universality of our simulated halo catalogues and illustrate its physical origin by comparing  haloes across simulations. In \S\ref{sec:modelling} we model the departures from universality as a function of an effective growth rate and power spectrum slope in each cosmological model. In \S\ref{sec:validation} we validate these predictions against the halo mass function as measured in a suite of simulations spanning a broad range of cosmological parameters. We further show in \S\ref{sec:scaling} that our proposed model can be employed to improve the accuracy of cosmology-rescaling techniques. Finally, we conclude and summarise our findings in \S\ref{sec:conclusions}.

\section{Numerical Simulations}
\label{sec:simulations}

In this section we will describe our set of cosmological simulations and our measurements of the halo mass function. Specifically, in \S\ref{sec:cosmologies} we describe the cosmological models we consider and in \S\ref{sec:Nbody-simulations} the numerical setup of the respective simulations. In \S\ref{sec:massfunction} we discuss our measurements of the halo mass function, and how we account for numerical and discretization errors.

\subsection{Cosmological Models}
\label{sec:cosmologies}

\begin{figure}
\centering
\includegraphics[width=\columnwidth]{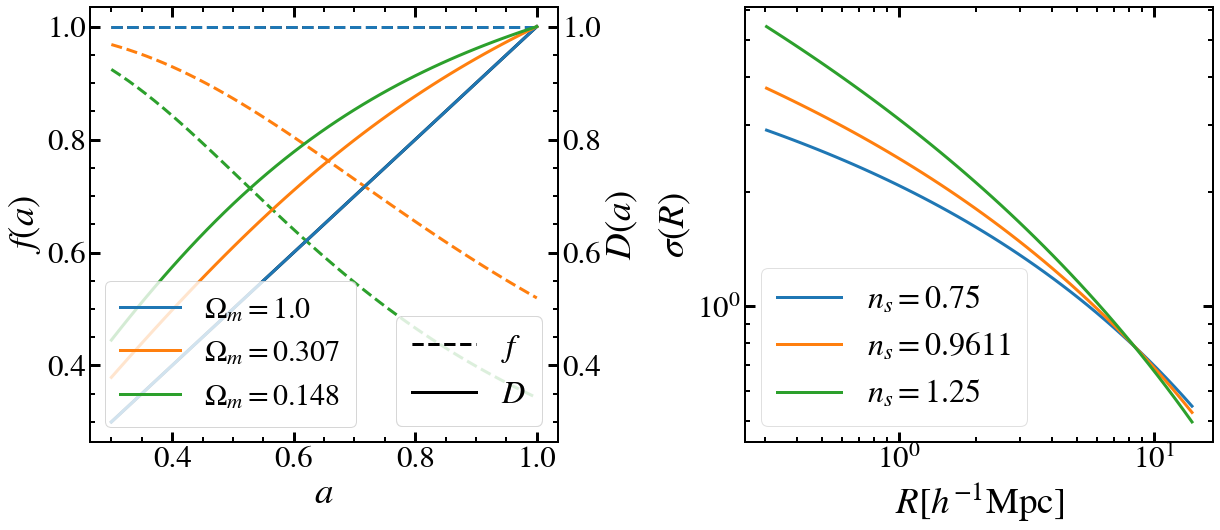}
\caption{\label{fig:linear} Linear properties of the cosmological models we consider and simulate throughout this paper. \textit{Left panel:} Growth factor, $D(a)$ and growth rate, $f(a)$, as a function of the expansion factor $a$. \textit{Right panel:} Linear mass variance at $z=0$ as a function of the Lagrangian radius of haloes of mass $M$.}
\end{figure}

We will consider $9$ cosmological models given by a combination of $3$ different growth histories and $3$ linear power spectra. In this way we can explore the effect of the growth history at a fixed linear mass variance, and of the power spectrum shape at a fixed growth history. We note that, in practice, we obtain varying growth histories by defining them with different values of the matter density parameter, $\Omega_m$, and vary the power spectrum shape by considering different values of the primordial spectral index $n_s$ (see Table~\ref{tab:cosmologies}).

In the left panel of Figure~\ref{fig:linear} we show the linear growth factor, $D(a)$, and growth rate, $f \equiv \frac{d \log D}{d \log a}$, as a function of expansion factor $a$ of the models we will consider. By construction, at $z=0$ all models have the same linear amplitude, however, they show very different values for the linear growth rate. At one extreme (green lines) we have a cosmology where structure initially grew very quickly and then stalled, where we expect very little mass accretion today. At the other extreme  (blue lines) is a cosmology where structure has been growing at the same pace through the history of the Universe, and in particular, we expect it to yield the highest present-day accretion rates onto dark matter haloes.

In the right panel of Figure~\ref{fig:linear} we show the 3 different $\sigma(R)$ at $z=0$ we consider. The respective power spectra are given by linear predictions for a cosmology consistent with recent observational constraints  (c.f. Table~\ref{tab:cosmologies}), for three different values of the primordial spectral index, $n_s = \{0.75, 0.96, 1.25 \}$. Although these values are clearly inconsistent with current data, they will allow us to clearly identify the role of the shape of fluctuations at a fixed growth history. Specifically, the cosmology with $n_s=0.75$ displays a very flat power spectrum, thus the density field has more similar fluctuations on all scales. We expect this to yield to similar collapse redshifts among different halo masses. On the other hand, the case with $n_s=1.25$ features stronger small scales fluctuations, thus we expect small haloes collapsing at high redshifts and large haloes forming at progressively later time.

\subsection{$N$-body simulations}
\label{sec:Nbody-simulations}

For each cosmological model described in the previous subsection, we have carried out a suite of cosmological simulations with $N=1024^3$ particles and $4$ different box sizes, $L = \{200,600,1200, 2400\}\,\hMpc$. This allows us to compute the halo mass function over a broad range of halo masses with a sufficient statistical accuracy at a moderate computational cost. Therefore, in total we have a suite of $36$ simulations. The details of the simulations are listed in Table~\ref{tab:sims}.

Each of our simulations is initialized at $z=49$ using second-order Lagrangian perturbation theory. As recently pointed out by \cite{Michaux:2020}, this configuration is expected to be accurate at the 2\% level for the abundance haloes resolved with more than $100$ particles.

We carry out our simulations with an updated version of the {\tt L-Gadget3} code \citep{Angulo:2020}, employing 48 MPI Tasks. In all cases, we set the Plummer-equivalent softening length to a $2\%$ of the mean interparticle separation. Each of our simulations took approximately $1$ to $3$ thousand CPU hours, depending on the mass resolution of the simulation.

\begin{table}
  \centering
   \caption{\label{tab:cosmologies} The cosmological parameters that we vary to obtain the $9$ cosmological models we simulate. We keep the rest of the cosmological parameters fixed assuming flat cosmology and $\Omega_b=0.046$, $\sigma_8=0.82$, $h=0.677$, $\Omega_{\nu}= 0.$, $w_0 = 0.0$, $w_a= 0.0$. }
  \begin{tabular}{l|r r l}
  	& $\rm extreme1 $  & $\rm central$  & $\rm extreme2$ \\
   \hline \hline
   \multirow{3}{*}{$n_{\rm s}$} & 0.75  & 0.75 & 0.75 \\
    & 0.9611  & 0.9611 & 0.9611 \\
    & 1.25 & 1.25 & 1.25 \\ \hline
    $\Omega_{\rm m}$ & 1. & 0.307 & 0.148 \\
    $\Omega_{\rm \Lambda}$ & 0.  & 0.693  & 0.852 \\ \hline
    %$\Omega_{\rm b}$  & 0.046 & 0.046 & 0.046 \\
    %$\Omega_{\nu}$  & 0 &0 & 0 \\
    %$\sigma_8$ & 0.82 & 0.82 & 0.82 \\
    %$h$ & 0.677  & 0.677 & 0.677 \\
    %$\omega_0$  & -1 & -1 & -1 \\
    %$\omega_{\rm a}$  &0 &0 &0  \\ \hline
       \end{tabular}
\end{table}

\begin{table}
  \centering
    \caption{\label{tab:sims} The main numerical parameters of our simulations. $L$ is the box size; $\epsilon$ the gravitational softening length; and $m_{\rm p}$ the mass of each $N$-body particle.}
   \resizebox{0.97\columnwidth}{!}{
    \begin{tabular}{l l | l l l l l l}
   \hline
  $L[\hMpc]$ & & 200 & 600 & 1200 & 2400 \\ \hline
  $\epsilon[\hMpc]$ & &  0.004 & 0.012 & 0.023 & 0.047 \\ \hline
  \multirow{3}{*}{$m_{\rm p}[\hMsun/1e10]$} & $\rm extreme1$ &0.207 & 5.58 & 44.67 & 357.3\\
							   & $\rm central$ &0.064 & 1.71 & 13.7 &  109.7 \\
							   & $\rm extreme2$  & 0.031 & 0.82 & 6.62 & 52.9 \\
\hline
   \end{tabular}}
\end{table}

\subsection{Halo Catalogues and discreteness correction}
\label{sec:massfunction}

\begin{figure}
\centering % \begin{center}/\end{center} takes some additional vertical space
\includegraphics[width=\columnwidth]{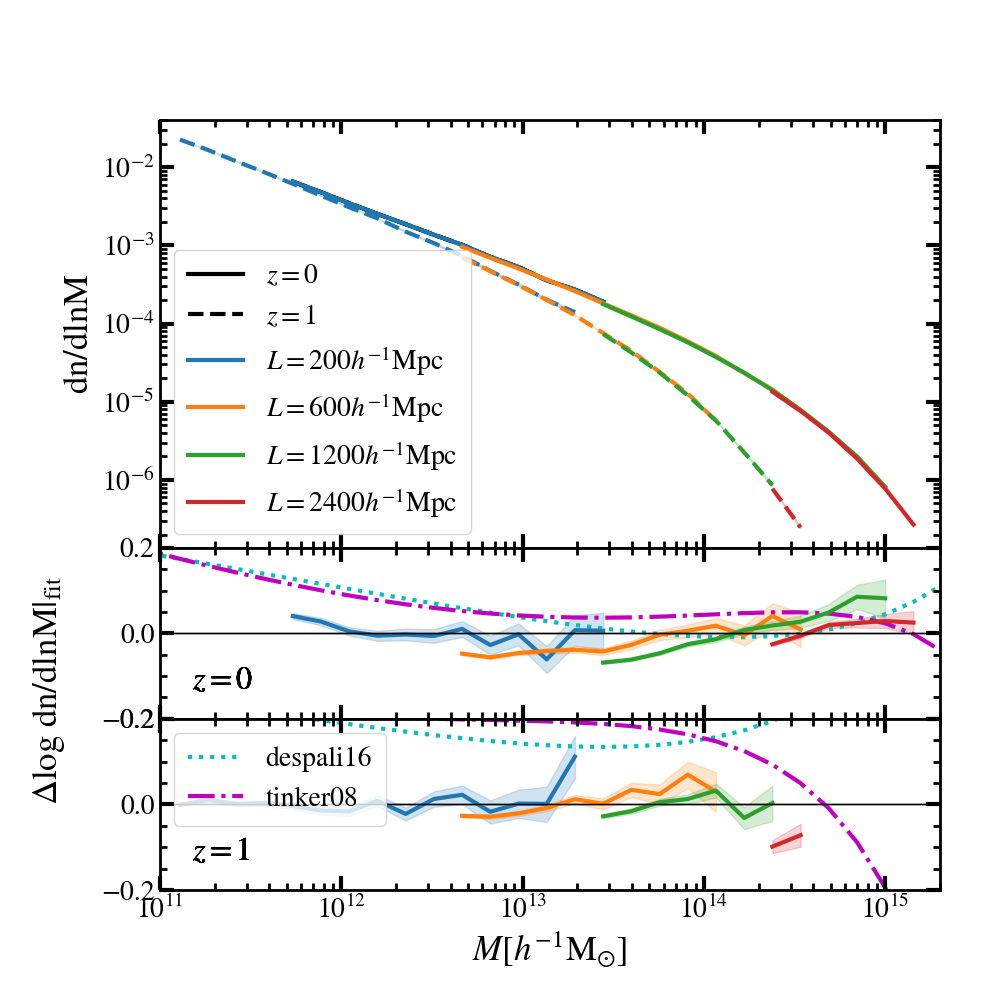}
\caption{\label{fig:example_massfunction_bestfit} The differential abundance of $M_{\rm 200b}$ haloes in one of our cosmological model ($\Omega_{\rm m}=0.307$, $n_{\rm s}=0.9611$) at $z=0$ and $z=1$, as estimated in 4 cosmological simulations of various sizes. The middle and bottom panels display the measurements relative to the expectations of the fitting function with dependence on $\neff$ and $\aeff$ developed in this work. The shaded regions correspond to the Poisson uncertainty of the measurements. Moreover, for comparison, dotted lines display the fitting functions developed in \protect\cite{Despali:2015} and \protect\cite{Tinker:2008}, as indicated by the legend.}
\end{figure}

We construct halo catalogues employing a Friends-of-Friends (FoF) algorithm with a linking length parameter $b=0.2$ at the $z=\{0,\,0.5,\,1\}$ simulation outputs. Additionally, for each FoF halo we compute the spherical-overdensity masses $M_{\Delta} = \frac{4\pi}{3}\Delta r_{\Delta}^{3}$, for $\Delta = \{200 \rho_c, 200 \rho_b, \Delta_{\rm vir} \}$, where $\rho_b$ is the mean matter density of the Universe, and $\Delta_{\rm vir} \equiv \rho_c \{18\pi^2 - 82[1-\Omega_m(z)] - 39[1-\Omega_m(z)]^2\}$ is the virial overdensity expected at each cosmological model. For each mass definition, we compute the halo mass function by considering haloes with more than $32$ particles in equally-spaced logarithmic bins, $\Delta \log M = 0.155$.

It is known that the FoF algorithm suffers from effects related to particle discreteness, which leads to an overestimation of the mass function \citep[see e.g.][and references therein]{Leroy:2021}. In \cite{Warren:2006} an empirical formula was derived to correct for these effects. In agreement with \cite{Lukic:2009} and \cite{more:2011}, we have, however, found that the performance of this correction varied greatly with cosmology and redshift. In addition, finite numerical precision in the force calculation and time integration, as well as the effect of softening length, also affect the abundance of haloes detected by FoF \citep{Ludlow:2019}. Thus, we have followed a conservative approach and impose a cut of 200 particles per halo without any additional correction. This limit, as shown by \cite{Ludlow:2019} is enough to keep all the numerical effects below $5\%$ for all mass definitions. %\todo{discuss} \cite{garrison:2019}
Consequently, we will add in quadrature to Poisson errors this 5\% to account for possible systematic errors in the measurement of our mass functions.

\subsection{Finite volume and output redshift corrections}

In order to span a broad mass range we have combined simulations of many box sizes. For different box sizes, however, the output times can vary slightly since in {\tt L-Gagdet3} we choose them to coincide with a global timesteps which, in turn, can vary from simulation to simulation. In Appendix \ref{sec:redsh_corr} we describe and validate a simple model with which we account for this effect in our measurements.

In addition, the lack of modes larger than the simulated box will induce systematic differences among different box sizes \citep[e.g.][]{Power:2006,Lukic:2007,Reed:2007}. However, we have checked that for all the boxes these effects are sub-percent at the relevant masses.

In Figure~\ref{fig:example_massfunction_bestfit} we display our measured $M_{\rm 200b}$ halo mass function at $z=0$ for the cosmological model with $\Omega_{\rm m}=0.307$ and $n_s=0.9611$. Results from simulations of various box sizes (after the corrections described above) are denoted by different line colours, as indicated by the legend.

The top panel displays the differential mass function whereas the middle and bottom panels display the ratio with respect to the predictions of a fitting function we will develop later in this work. Note we only display bins with more than $400$ objects resolved with at least $200$ particles. We can see how our suite of various box sizes complement each other to cover a very large range of halo masses, from $5 \times 10^{11}\hMsun$ up to $10^{15}\hMsun$. The agreement in the overlapping regions is always better than 5\%, consistent with our systematic error estimate. Although not shown here, we have checked that this also holds for the other 8 cosmological models.

For comparison, in the bottom panels we also display the fitting functions of \cite{Despali:2015} and \cite{Tinker:2008}. Although some differences among our data and these models are expected due to differences in the group finder, the comparison readily highlights the impact of non-universality of the mass function. At $z=0$ our model and that of \cite{Despali:2015} and \cite{Tinker:2008} are in reasonable agreement. However, at $z=1$, these fits overestimate by more than 10-15\% the abundance of haloes in our simulations. In subsequent sections we will explore this issue in greater detail.

\section{The dependence of the mass function on growth history}
\label{sec:nonuniversality}

In this section we will compare how the same linear fluctuation turns into collapsed objects of different mass for different cosmologies. We will then explore the dependence of the mass function on both growth rate and the slope of the power spectrum.

\subsection{Examples of haloes matched across simulations}\label{subsection:crossmatching}
\begin{figure}
\begin{tabular}{cc}
  \includegraphics[width=0.47\columnwidth]{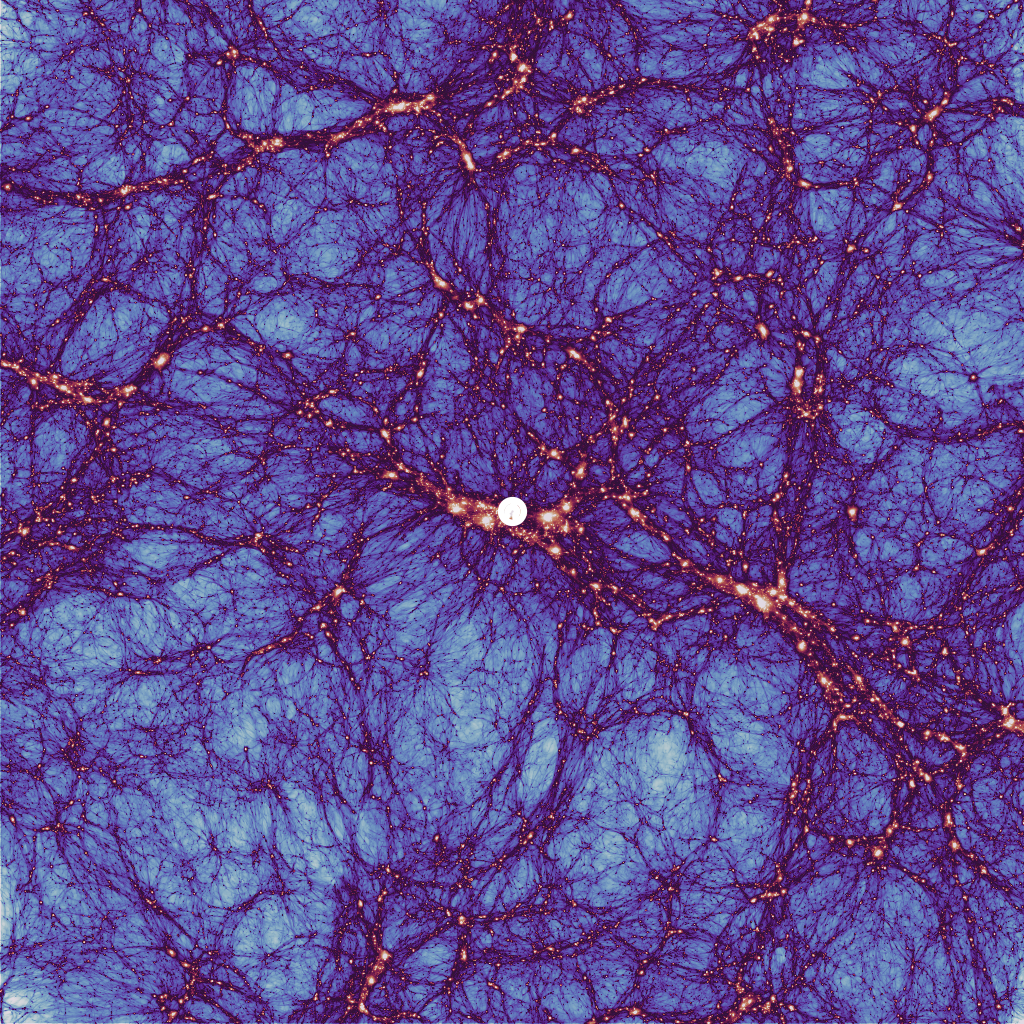} &   \includegraphics[width=0.47\columnwidth]{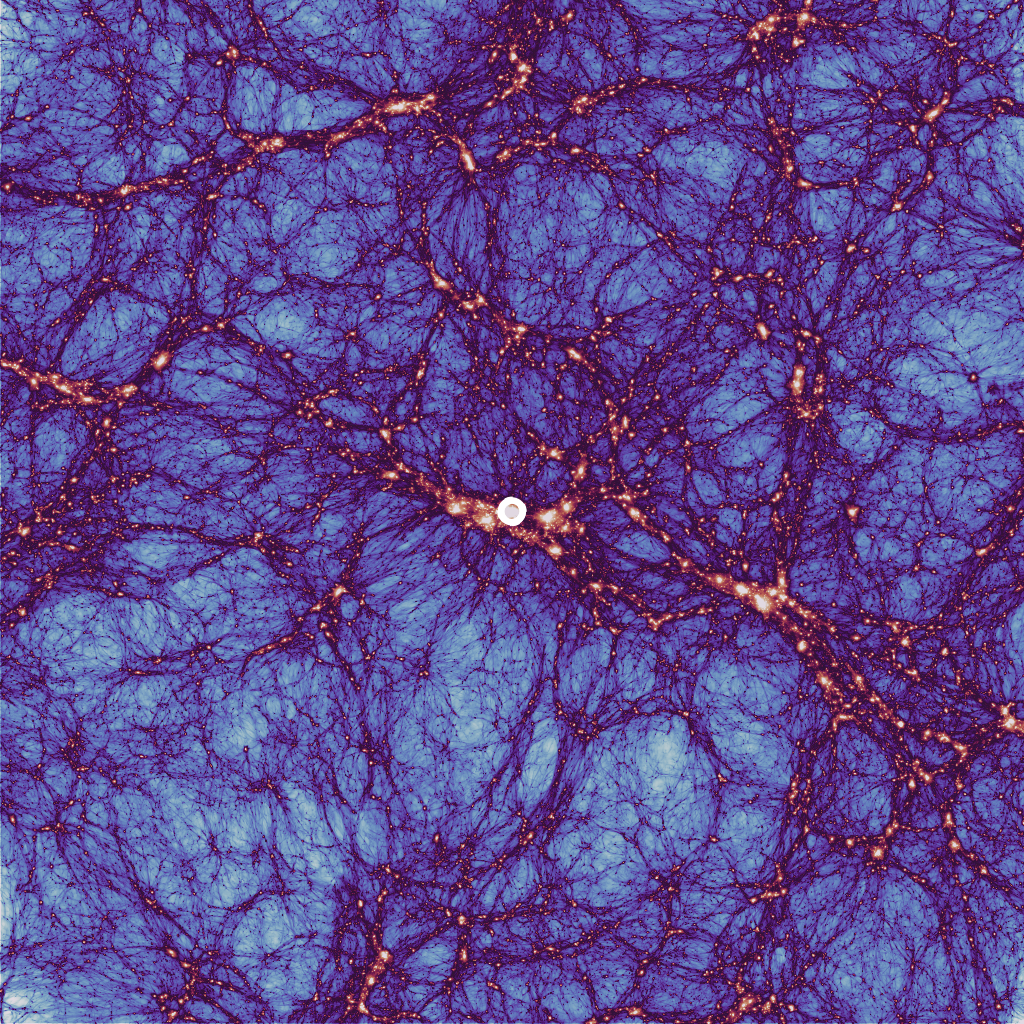} \\
%(a) first & (b) second \\[6pt]
 \includegraphics[width=0.47\columnwidth]{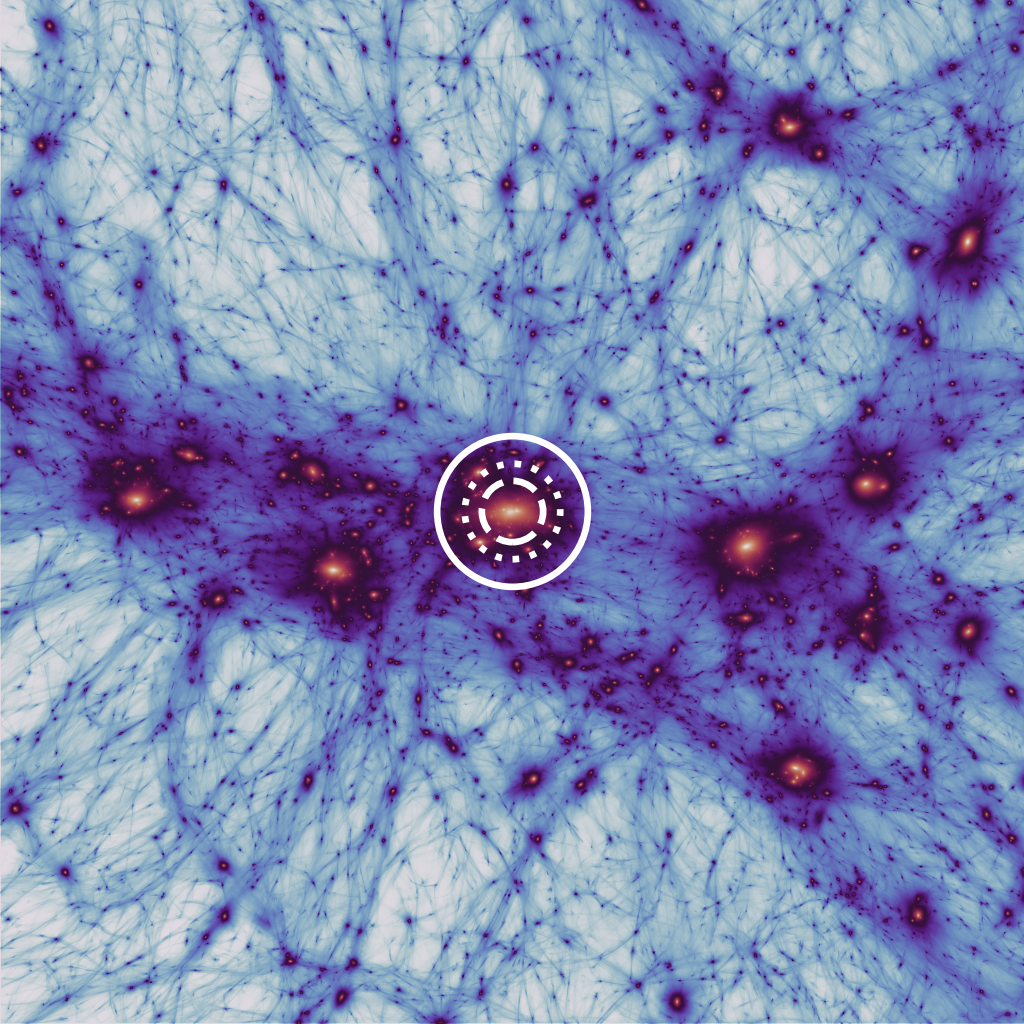} &   \includegraphics[width=0.47\columnwidth]{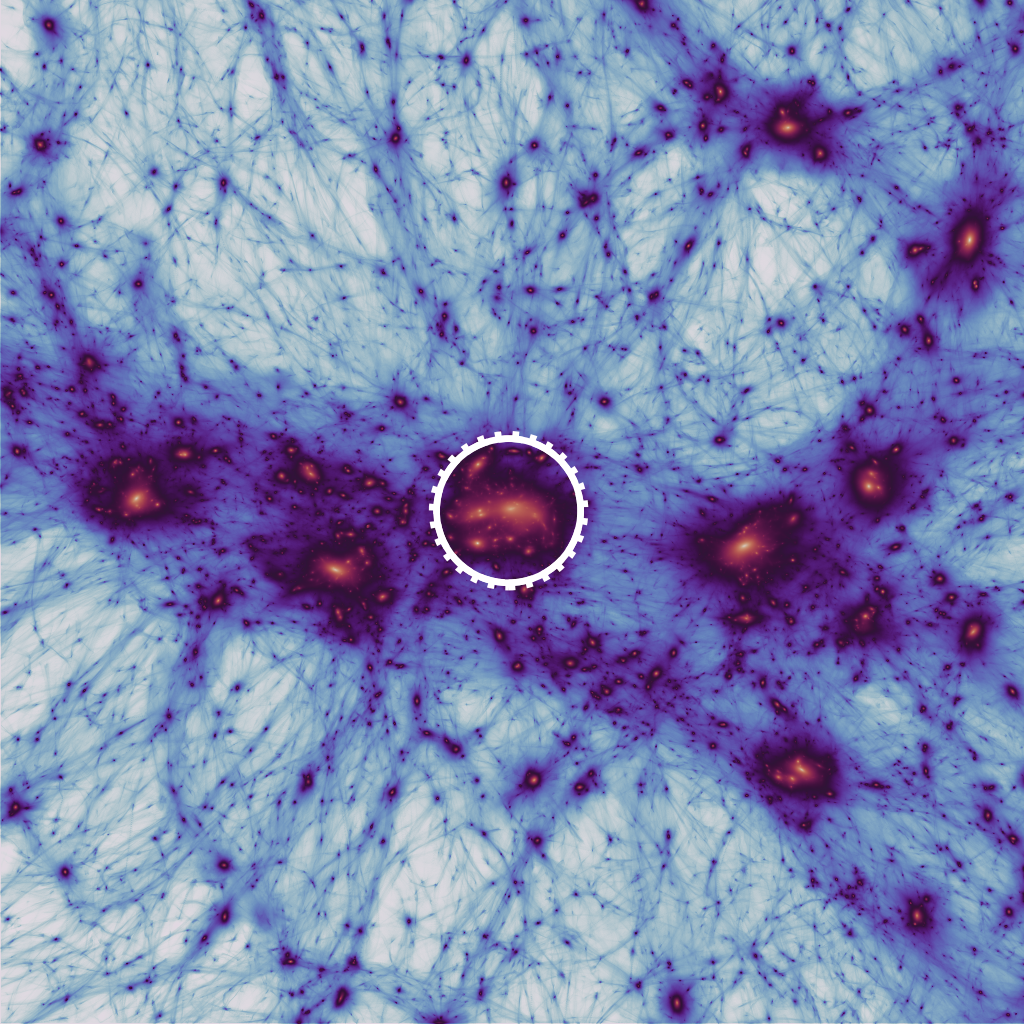} \\
%(c) third & (d) fourth \\[6pt]
\includegraphics[width=0.47\columnwidth]{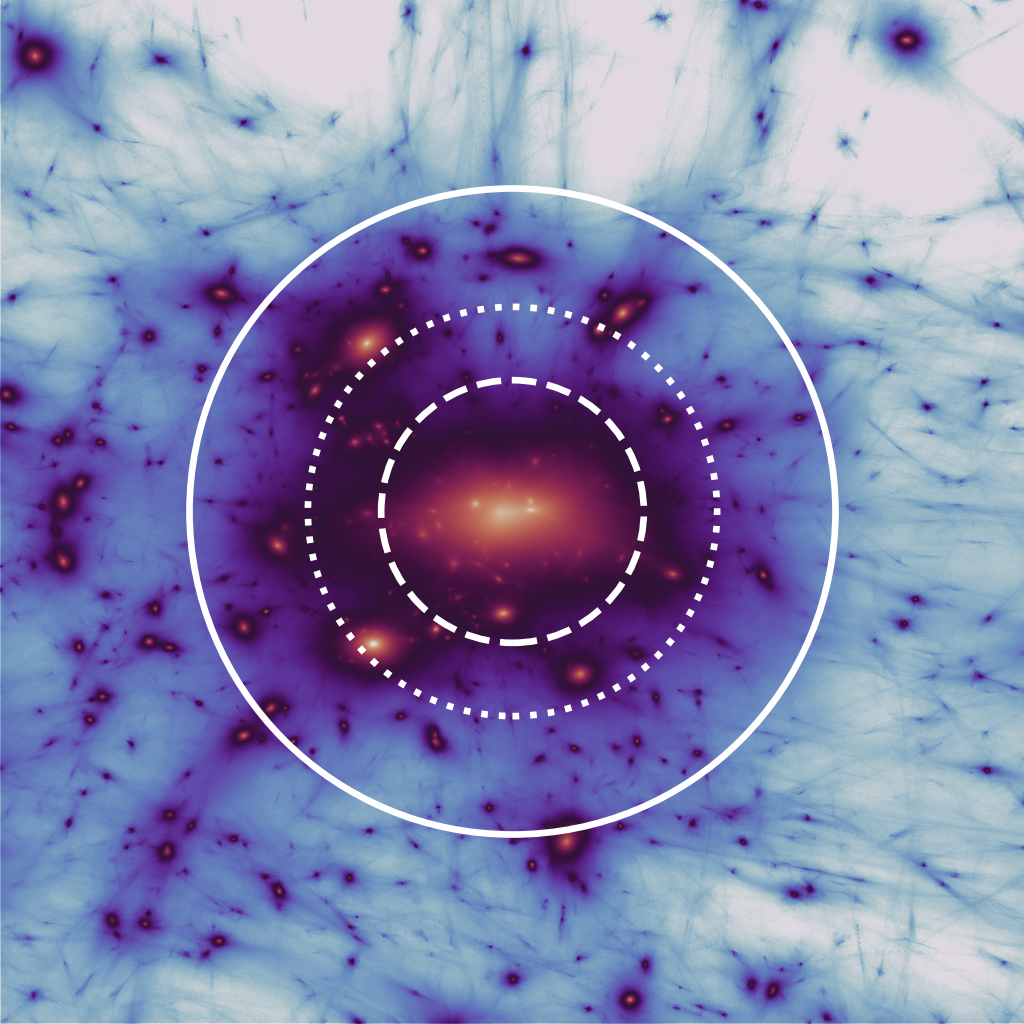} &   \includegraphics[width=0.47\columnwidth]{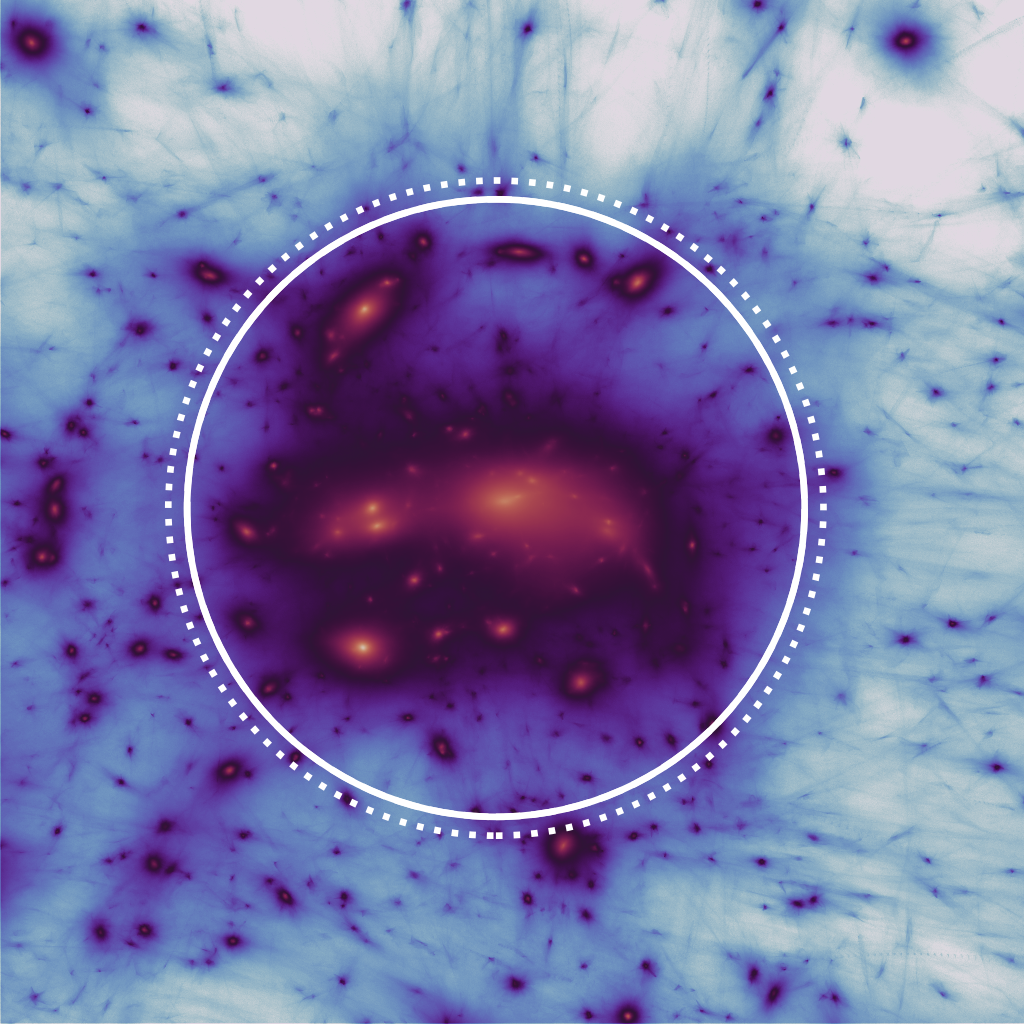} \\
\end{tabular}
\caption{\label{fig:fields} The projected simulated density field normalised by the mean background density at $z=0$ for two cosmological models that share the same linear density field but differ significantly in their current growth rate. Top panels show the full simulated box, L=$200\hMpc$, whereas the middle panels and bottom panel show zooms into regions of $30\hMpc$ and $7\hMpc$ a side centered in a halo of normalised mass $M_{\rm 200b} / \rho_{\rm b} \sim 8.5 \times 10^{3} h \rm Mpc^{3}$. In the left column we plot the cosmological simulation with the lowest matter density $\Omega_m=0.148$, whereas in the right panel we display that with the highest matter density, $\Omega_m=1$. The solid, dashed and dotted circles show $r_{\rm 200_b}$, $r_{\rm 200_c}$ and $r_{\rm vir}$ radii of the halo.}
\end{figure}

%-----------------------------------------
\begin{figure}
\centering
\includegraphics[width=\columnwidth]{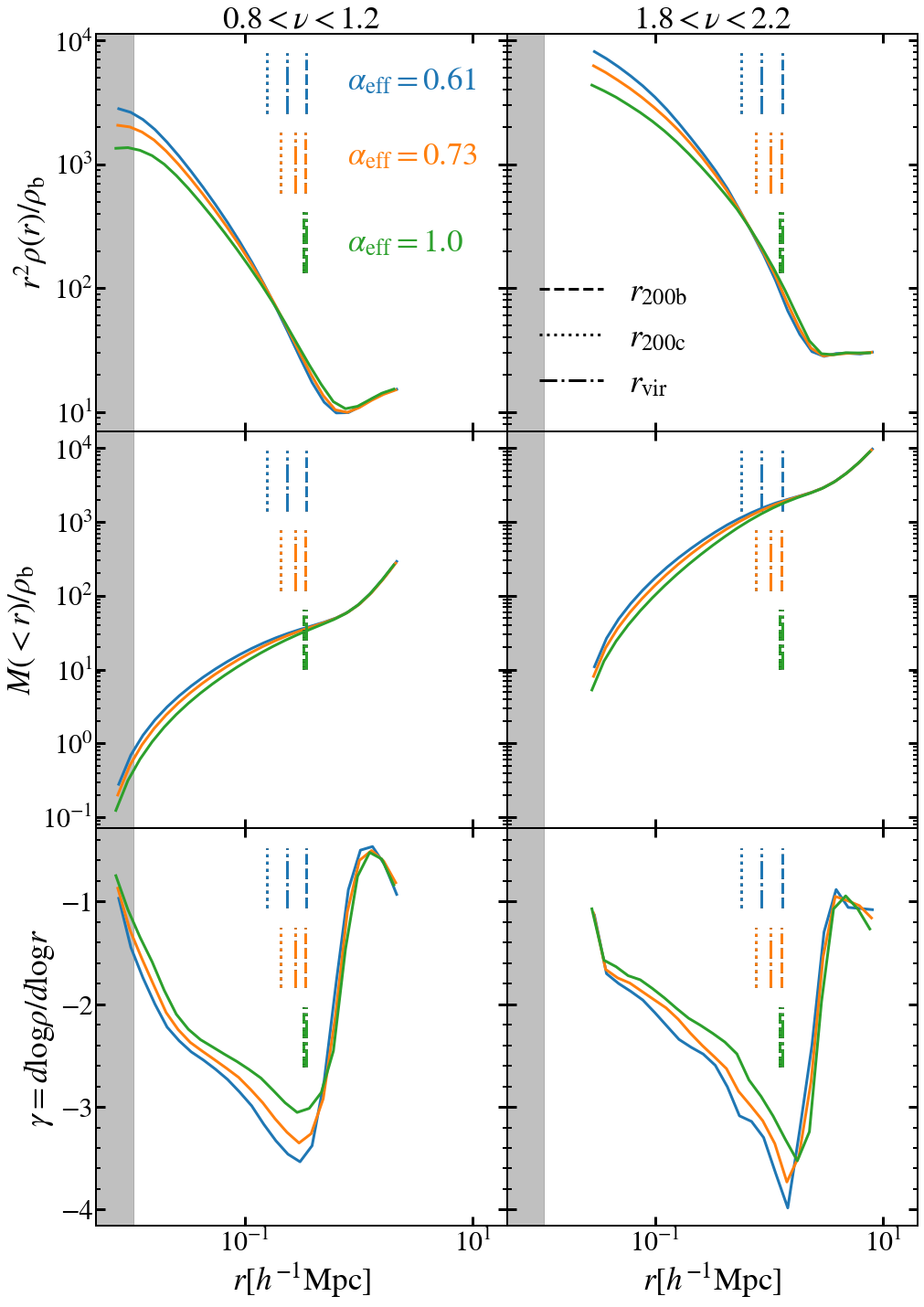}
\hfill
\caption{\label{fig:matched} The profiles of the crossmatched haloes with the same linear density field for two $\nu$ bins. In the first row we display the density profiles, in the second row the cumulative mass and in the third row the logarithmic slope of the density profile. The vertical lines indicate the values of $r_{\rm 200c}$, $r_{\rm vir}$  and $r_{\rm 200m}$ radii. The shaded area represents $r < 2.7 \epsilon$, where $\epsilon$ is the softening length.}
\end{figure}

\begin{figure}
\centering
\includegraphics[width=\columnwidth]{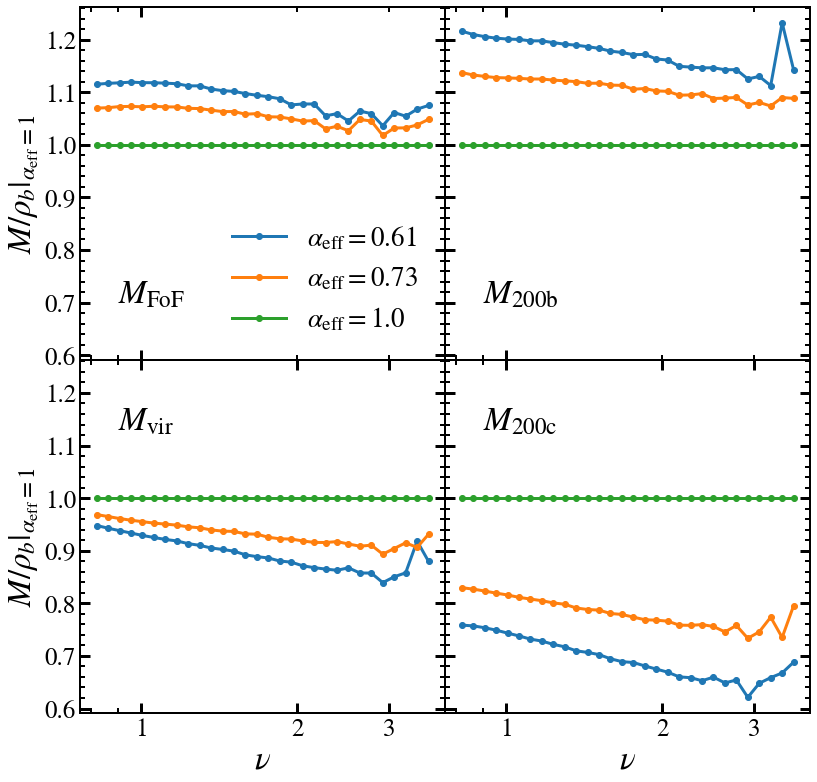}
\hfill
\caption{\label{fig:matched_mass_aeff} The mass ratio of the cross-matched haloes at $z=0$ respect to the self-similar cosmology, i.e. $\aeff = 1$. The colours depict the current growth rate value of the given cosmology. In each panel we display the ratios for $M_{\rm FoF}$, $M_{\rm 200b}$, $M_{\rm vir}$, and $M_{\rm 200c}$ mass definitions.}
\end{figure}

\begin{figure}
\centering
\includegraphics[width=\columnwidth]{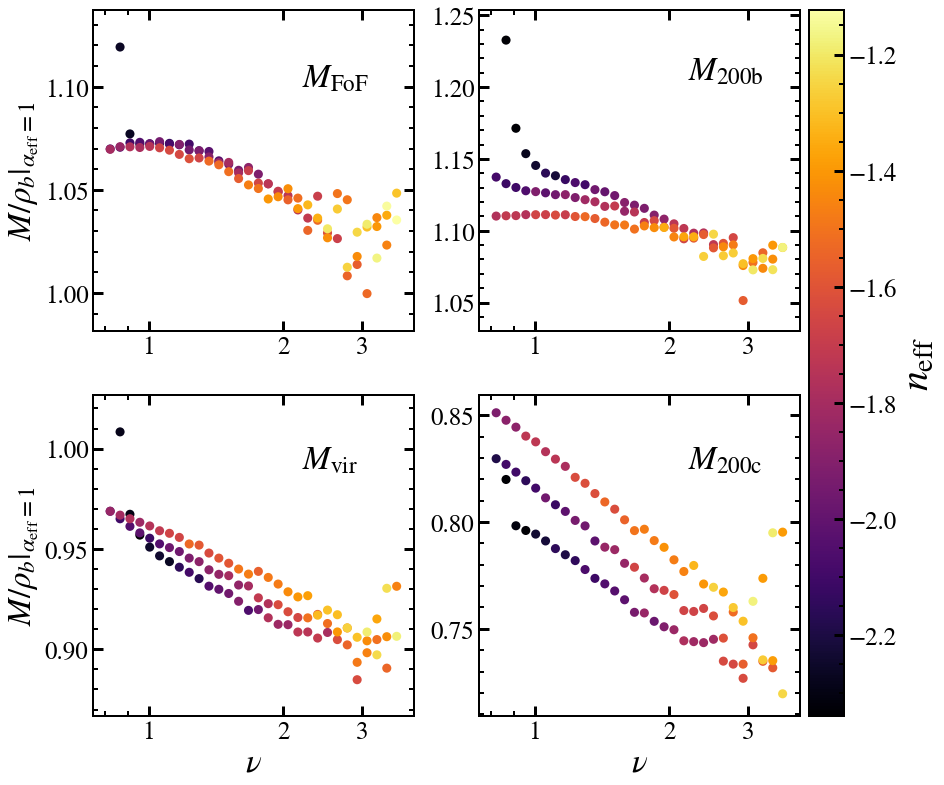}
\hfill
\caption{\label{fig:matched_mass_neff} The mass ratio of the cross-matched haloes at $z=0$ with respect to the cosmological model with self-similar growth, i.e. $\aeff = 1$, and identical linear density field. The colours represent the $\neff$ values of the cross-matched haloes. In each panel we display the ratios for $M_{\rm FoF}$, $M_{\rm 200b}$, $M_{\rm vir}$, and $M_{\rm 200c}$ mass definitions.}
\end{figure}
%-------------------

The universality of the mass functions assumes that the mass function is completely described by the linear density field. In order to test this assumption, we have run simulations with very different growth histories that share the same linear density field at $z=0$. In Figure~\ref{fig:fields} we show the simulated density field at $z=0$ for our $L=200\hMpc$ simulations with the most dissimilar growth histories for the $n_s=0.9611$ cosmology. In the top panel we show a region of $200\hMpc$ wide, whereas in the middle and bottom panels we zoom on a massive dark matter halo of normalised mass $M_{\rm 200b} / \rho_{\rm b} \sim 8.5 \times 10^{3}  h \rm Mpc^{3}$. We can see that, although both cases corresponding to identical $z=0$ linear density peaks, their nonlinear counterparts are different. Specifically, in the case with the highest $\Omega_m$ value, and thus, highest current growth rate (in the right), haloes are significantly less dense in its center, which is consistent with its expected lower formation redshift and thus lower concentration parameters. Various definitions of halo radii are displayed by white circles in each case. By comparing $r_{\rm 200c}$ and $r_{\rm vir}$ radii in the two cosmologies, we see that they identify very different regions of the halo, unlike $r_{\rm 200b}$ which is similar in both cases. %We will explore this further in \S\ref{subsection:crossmatching}.

To explore this further, we will compare haloes of the same peak height in different simulations. In particular, we have cross-matched halo catalogues among simulations that share the same linear power spectrum at $z=0$. For this, we associate two haloes based on their position and peak height. From Figure \ref{fig:fields} we expect that the same fluctuation in the linear density field will end up having a different mass depending on its nonlinear evolution.

We will characterise each halo by an "effective growth rate" and an effective "local power spectrum slope", which we define respectively as:

\begin{equation}
\aeff(a) \equiv \frac{d \log(D)}{d \log a}|_{a=a_{ev}},
\end{equation}

\noindent where $a_{\rm ev}$ is defined implicitly via $D(a_{\rm ev}) = \gamma D(a)$ with $\gamma=4/5$, and

\begin{equation}
n_{\rm eff} \equiv -3 - 2 \frac{d \log \sigma(R)}{d \log R}|_{\kappa R_L(M)}
\end{equation}

\noindent where $\kappa=1$, and $R_{L}$ is the Lagrangian radius of a halo of mass $M$. Physically, these two parameters will be capturing how quickly haloes have recently grown and the density profile of the collapsing region, which can be considered as a proxy for the full mass accretion and merger history of a given halo.

Note that the effective growth rate is not evaluated at the redshift in which we identify a halo, but it is evaluated in the past, i.e. $\gamma < 1$. By this we seek to capture not the rate of current mass accretion, but instead the amount of mass that has been accreted recently. We have tried different definitions of $\aeff$ and found that this distinction was particularly important for models with dynamical dark energy. We chose the numerical value for $\gamma$ as that which provided the most accurate and simplest model for the halo mass functions, as we will show in Section \ref{sec:scaling}.

In Figure~\ref{fig:matched} we display the spherically averaged mass distribution around crossmatched haloes in 2 bins of the peak height, $\nu \sim1$ and $\sim2$. We display the average density profile, the cumulative mass profile, and the logarithmic slope of the halo density profiles. Vertical lines indicate the radius at which the average enclosed density reaches a value equal to 200 times the background, virial and critical density, as indicated in the legend. Coloured lines indicate three growth histories for the simulations with $n_s = 0.9611$.

We can see that generically the mass profiles differ systematically with $\aeff$, at all values of $\nu$. The higher the growth rate the lower the  enclosed mass respect to the background density at a given physical radius. We emphasise that all these objects share the same shape and amplitude of their linear overdensity field at $z=0$. Thus all changes necessarily are caused by the different growth history.

The different growth histories are expected to influence the internal structure of haloes. In particular, lower growth rates are expected to cause lower current accretion rates onto haloes, which implies haloes formed earlier and thus are expected to have higher concentrations. In the first row of panels we see that this is indeed the case. Inner regions of haloes appear more concentrated. However, the changes are not limited to the concentration, as external parts are also modified increasing their density the higher the current growth rate. In fact, there seems to be an inflection point located at around $r_{\rm 200c}$ radius.

Despite the systematic dependence on $\aeff$, the profiles are very similar when expressed in physical units. However, as a consequence of the pseudo-evolution of the halo boundaries \citep{Diemer:2013}, when expressed in $r_{\Delta}$ units the profiles become very different. The pseudo-evolution of the boundaries is clear in Figure~\ref{fig:matched}. In the $\aeff=1$ cosmology $r_{\rm 200c}$ is almost three times larger than in the $\aeff=0.343$ case, while $r_{\rm 200b}$ radii remain roughly constant. Thus, depending on how we define the boundary of our halo, the mass differences will be enhanced or suppressed.

In the lower panels of Figure~\ref{fig:fields} this can be appreciated visually. The panels show the most massive crossmatched halo at $z=0$ in $\aeff=0.343$ (left) and $\aeff=1$ (right) cosmologies. The dashed, dotteed and solid lines represent $r_{\rm 200c}$, $r_{\rm vir}$ and $r_{\rm 200b}$ radii of the halo. While $r_{\rm 200b}$  defines a halo boundary roughly at the same physical location, $r_{\rm 200c}$ compares very different regions of the density field. This effect is less important for $r_{\rm vir}$, which might explain why \cite{Despali:2015} found that $M_{\rm vir}$ is the mass definition that leads to the most universal behavior.

In order to explore this effect more systematically, in Figure~\ref{fig:matched_mass_aeff} we show how the masses of the crossmatched haloes differ depending on the growth history and the mass definition. We display the ratio of the masses of the crossmatched haloes respect to the $\aeff=1$ cosmology as a function of $\nu$. We see that in the cosmology with the lowest current growth rate value $M_{\rm 200c}$ masses are around $30\%$ smaller than in our reference cosmology. However, for the same cosmologies and haloes, $M_{\rm 200b}$ masses are around $20\%$ more massive.

This effect has two contributions. On the one hand, $r_{\rm 200c}$ radii lie in the inner parts where the effect of the growth history on the mass profile is larger. On the other hand, because of the pseudo evolution of $r_{\rm 200c}$, we compare the masses enclosed in different physical radii. As a consequence, even if at a given physical radius the enclosed mass is always larger for haloes in low growth rate cosmologies, when comparing $M_{\rm 200c}$ masses it seems that haloes in high growth rate cosmologies are more massive. Note that this is solely because we compare masses enclosed in different physical regions. Thus, the non-universality of $M_{\rm 200c}$ mass function is in a big part due to the evolution of the boundary itself.

Finally, we want to explore the effect of the local slope of the power spectrum in the crossmatched haloes. For a given power spectrum, redshift and $\nu$, $\neff$ is completely determined. Therefore, in order to see the effect of this variable on the mass of the haloes, we crossmatch the cosmologies with $\aeff=0.52$ and $\aeff=1$ for the three power spectra defined with the three $n_s$ values we have considered in this work. Note that we only crossmatch cosmologies with the same linear power spectrum. However, if the local power spectrum slope affects the mass of the haloes, we expect the departures of the masses of $\aeff=0.52$  from $\aeff=1$ cosmology to be different in the three linear density fields. In Figure~\ref{fig:matched_mass_neff} we show the results following Figure~\ref{fig:matched_mass_aeff}, coloured by the $\neff$ values of the crossmatched haloes. Indeed, we see that for a given $\nu$, the departures of $\aeff=0.52$ haloes from $\aeff=1$ haloes are different depending on the $\neff$ value of the halo. Nevertheless, these differences are much smaller than the differences that haloes with different $\aeff$ show.

In summary, the whole density profile of the halo is affected by the growth history in a non trivial way. This effect will be reflected in the mass function in a different fashion depending on how masses are defined.  Specifically, we expect the non-universality of the mass function to change with the mass definition.

\subsection{The non-universality of the mass function}

From the numerical simulations described in the previous section, we have obtained measurements of the halo mass function in 9 cosmologies -- 3 growth histories and 3 different power spectrum slopes -- covering a broad range of masses at various redshifts.

In order to compare these measurements and estimate the impact of the non-universality in the mass function, we have computed $\nu\fnu$, where $\nu$ is $\delta_c/\sigma(M)$. Operationally, we first measure the mass function $n(M)$ and then estimate $\nu\fnu$ inverting Equation \ref{eq:mass_function}.

The critical density for collapse, $\delta_c(z)$, can be estimated as $3/5 (1.5 \pi)^{2/3} \Omega_{\rm m}(z)^{0.005}$ \citep{Kitayama:1996}, with an explicit dependence on $\Omega_{\rm m}$. In this work we approximate it with the value that corresponds to a universe where there is only matter, $\delta_c=1.686$. We do this to simplify the redshift and cosmology dependence and make it easier to model. We have checked that the deviations from universal behaviour of the mass function are stronger (up to $\pm 30\%$) than the effect of taking into account the redshift evolution of the critical overdensity for collapse, which changes the results around $\pm 10\%$ (see Appendix \ref{sec:crit_z}).

%-----------------------------------------
\begin{figure}
\centering
\includegraphics[width=\columnwidth]{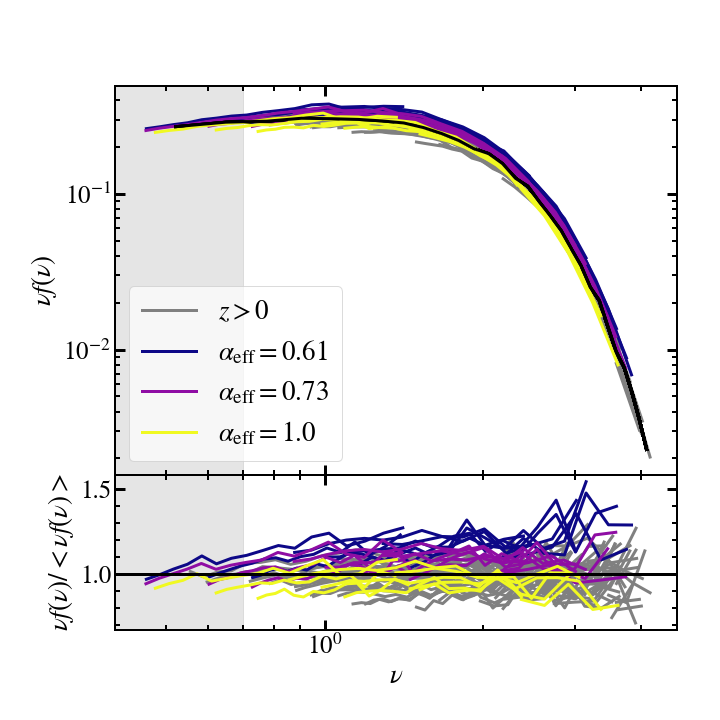}
\hfill
\caption{\label{fig:mass_functions} The measured $M_{\rm 200b}$ mass functions of the 9 cosmological models considered in this work at $z=0$, $z=0.5$ and $z=1$. The bottom panel shows the ratio relative to the mean value in each $\nu$ bin. We show in gray the $z>0$ mass functions, while we colour the $z=0$ mass functions according to their $\alpha_{\rm eff}$ value. The shaded area corresponds to measurements with $\nu < 0.7$, which will be excluded when developing a fitting function for $f(\nu)$.}
\end{figure}
%-------------------

In Figure~\ref{fig:mass_functions} we display the measurements of $\nu\fnu$ from $M_{\rm 200b}$ mass function in the cosmologies listed in the Table~\ref{tab:cosmologies} at $z=0$, $z=0.5$ and $z=1$. The mass functions with $z>0$ are displayed as gray lines, while $z=0$ mass functions are coloured according to their $\alpha_{\rm eff}$ value. The average value in each $\nu$ bin is displayed as a black solid line, and it is used as a reference for the ratio displayed in the bottom panel. To avoid possible biases due to differential coverage of our models \footnote{At a fixed volume and number of particles, the differences in the power spectrum shape and $\Omega_{\rm m}$ lead to differences in the range of $\nu$-peaks that our simulations are able to resolve.}, we will restrict our subsequent analysis to the range $0.7 < \nu < 5$. For the cosmology most consistent with observational constraint, this implies a mass range of $10^{11} < M/[\hMsun] < 10^{15}$ at $z=0$.

In this figure we can clearly see deviations from an universal behavior. For $\nu$ values above unity, haloes of a given peak height in one cosmology can be up to $70$\% more abundant than in others. By construction, the origin of this non-universal behavior must be in a combination of the different statistics of the initial Gaussian random fields and the different growth histories. Indeed, we can already see that this is the case for the cosmologies that share the same linear density field at $z=0$. At a given $\nu$, haloes seem to be more abundant the lower the growth rate value. Recall that, as we saw in the previous section, this is a consequence of the same fluctuation being more massive for low growth rate values.

\begin{figure*}
\centering % \begin{center}/\end{center} takes some additional vertical space
\includegraphics[width=\textwidth]{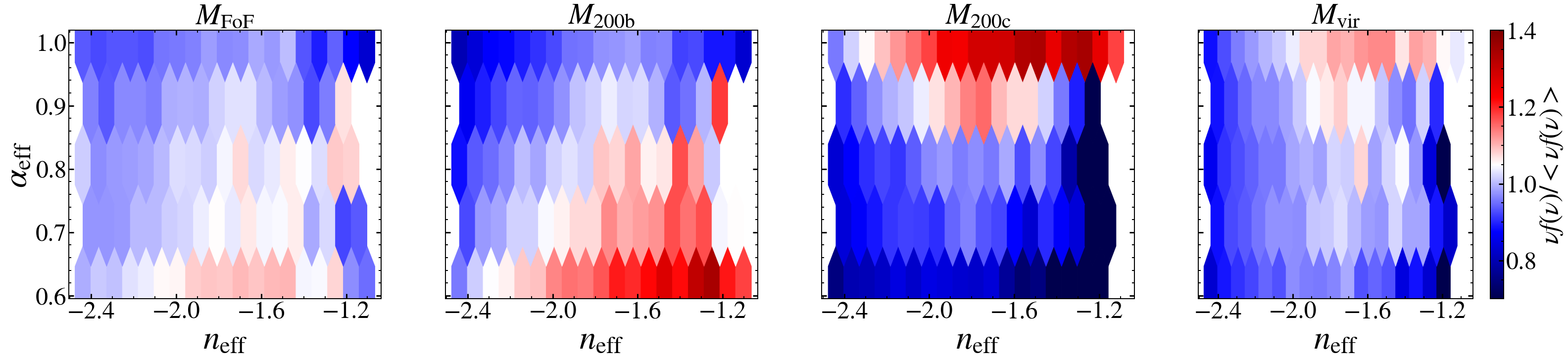}
\hfill
%\includegraphics[width=.45\textwidth]{}
% "\includegraphics" is very powerful; the graphicx package is already loaded
\caption{\label{fig:mass_function_delta} The deviation of the mass functions respect to the mean value (computed in each $\nu$ bin) plotted according to the $\aeff$ and $\neff$ values. The panels correspond to different mass definitions.}
\end{figure*}

We now explore how these deviations correlate the value of the effective growth rate and power spectrum slope at any redshift. In Figure~\ref{fig:mass_function_delta} we display the deviations from the mean $\nu\fnu$, at a fixed $\nu$ as a function of $\neff$ and $\aeff$ (horizontal and vertical axis). Each panel shows the result for a different mass definition.

For all mass definitions we can see that the non-universality clearly correlates with these properties -- despite them being only a proxy of very different merger and assembly histories. In the second panel we see that for $M_{\rm 200b}$, deviations around the mean can reach $\pm 20\%$. Cosmologies that have higher-than-the-mean $\nu\fnu$ typically have lower growth rate values, whereas those with higher growth rate values lead to lower abundances. At fixed $\aeff$, deviations from universality are much smaller, about $15\%$, and they correlate with $\neff$. Note that here we are plotting measurements of many redshifts, therefore, we expect that the redshift evolution of the mass function could be described through the dependence on these physically-motivated variables.

However, the non-universality of the mass function depends on the mass definition. Among those considered in this work, $M_{\rm 200c}$ ($M_{\rm FoF}$) mass functions are the most (least) non-universal with deviations up to $\pm 30\%$ ($\pm 10\%$) around the mean. It is interesting to note that, as \cite{Despali:2015} found, $M_{\rm vir}$ mass functions are the most universal among the SO mass functions. Furthermore, the dependence on the growth rate is inverted in $M_{\rm 200c}$ and $M_{\rm vir}$ cases with respect to $M_{\rm 200b}$ case \citep[in agreement with][]{Diemer:2020}.

In summary, haloes of a given peak height, $\nu_{\rm 200b}$, are more abundant the lower the growth rates and the shallower the power spectrum slope. In other words, a halo that forms early and has grown mostly trough minor mergers, will be more massive than another that has recently formed and has experienced a lot of major mergers, even if both have an identical peak height in the linearly extrapolated initial field.

There could be different paths to follow at this point. One could be to find the halo boundary definition that minimises the non-universality of the mass function. In fact, we have seen that mass definitions based on the critical density induce strong pseudo-evolution in the mass function driven by the change of the boundary of the halo. One quantity that has been argued is more physical is the turnaround radius, which by definition encloses the outermost shell that has collapsed. In the same direction, the first explorations of the splashback mass functions have been done \citep{Diemer:2020}. Other alternatives have been recently proposed, which are claimed separate better the linear and nonlinear regimes \citep[e.g.][]{Garcia:2020,Fong:2020}.

Any universal mass definition, as we saw earlier, would correspond to very large scales, which although perhaps better suited for describing the mass distribution, might not describe equally well, and thus it might display less correlation with the properties of collapsed gas and of the galaxies hosted by the halo. In addition, many of the proposed halo definitions are ambiguous to implement numerically.

Another option would be to develop a model for the changes of the full density profile as a function of the mass accretion history. For instance, in an analogous manner to the models developed for the relationship between the concentration and the expected mass accretion history in Extended Press Schechter \citep{Ludlow:2016,Ludlow:2019}, it is perhaps possible to develop a model for the outer regions of a halo, which would then predict the changes in halo mass at any radius.

The option we will follow here is to adopt a standard halo definition but calibrate the predictions for the halo abundance to be a function of the peak height but also of the properties of the cosmological model. We will show that with a simple parameterization in terms of $\aeff$ and $\neff$, we can accurately describe the halo mass function for a large region in cosmological parameters space.

\section{Modelling the dependence on growth function and power spectrum slope}\label{sec:modelling}

In the previous sections we showed how the halo mass function varies systematically with growth rate and slope of the power spectrum. In this section, we will model this dependence explicitly.

We will focus on the $M_{\rm 200b}$ mass function, because as it has been discussed in the previous Section, it is the most physically motivated choice and presents the least pseudo-evolution among the overdensity mass definitions. FoF mass function could also be a good candidate, but the fact that it has no clear observational counterpart and that the masses are very subject to numerical effects make it a less interesting candidate. However, in Appendix~\ref{sec:extension_to_masses} we show that our approach is valid to model the mass function of any of the other mass definitions considered in this work.

We have employed the following functional form $f(\nu,\neff,\aeff)$ to model each of our measurements

\begin{equation}\label{eq:x}
f(\nu, \neff, \aeff) = f_1(\nu)\,f_2(\neff)\,f_3(\aeff)
\end{equation}
\begin{equation}\label{eq:x_nu}
f_1(\nu) = 2A_{mp}\, (1 + (a\,\nu^{2})^{-p\,}) \sqrt{\frac{a\, \nu^{2}}{2\pi}} e^{- 0.5a\,\nu^{2}}
\end{equation}
\begin{equation}\label{eq:x_neff}
f_2(\neff) = n_0\,\neff^{2}+n_1\,\neff+n_2\,
\end{equation}
\begin{equation}\label{eq:x_aeff}
f_3(\aeff) = a_0\,\aeff+a_1\,
\end{equation}

\noindent where $\{a, p, A_{mp}, n_0, n_1, n_2, a_0, a_1\}$ are the free parameters of the model, where we have used the same functional form for $f(\nu)$ as \cite{Despali:2015}. Notice that the contributions of the variables are separable. Thus, in principle one could calibrate $f(\nu)$ separately, or reuse previously ran simulations.

In each case, we find the best fitting parameters by minimizing the $\chi^2$ of the quantity $\nu f(\nu)$. The uncertainty in each measurement is given by the Poisson statistics and a systematic uncertainty of $5$\%, as discussed in Section 2.  We impose a limit of $200$ particles per halo and $400$ haloes per mass bin. The minimization is done with the \texttt{optimize.minimize} package of \texttt{scipy}, imposing bounds on the possible values that the parameters may take. Other than our main model (Eq.~\ref{eq:x}), we have found the best fit parameters for the functional forms that depend only on $\nu$ (Eq.~\ref{eq:x_nu}), $\nu$ and $\neff$ (Eq.~\ref{eq:x_nu} $\times$ \ref{eq:x_neff}) and $\nu$ and $\aeff$ (Eq. ~\ref{eq:x_nu} $\times$ \ref{eq:x_aeff}). We list all the best-fit parameters in Table~\ref{tab:best_fit_pars_m200b}.

\begin{table*}
  \centering
  \caption{\label{tab:best_fit_pars_m200b} A table listing the best fit parameters of our $M_{200b}$ fitting functions.}
    \begin{tabular}{r r r r r r r r r}
  	& $a$ & $p$ & $A_{mp}$ & $n_0$ & $n_1$ & $n_2$ & $a_0$ & $a_1$\\
	   \hline
    $f_1(\nu)$ & 0.769 & 0.0722 & 0.3173 & -- & -- & -- & -- & -- \\
    $f_1(\nu)f_2(\neff)$ & 0.7741 & 0.1746 & 0.3038 & -0.1912 & -0.4211 & 0.9859 & -- & -- \\
    $f_1(\nu)f_3(\aeff)$ &0.772& 0.0308 & 0.3069 & -- &-- &-- & -0.6255 & 1.5654 \\
    $f(\nu, \neff, \aeff)$ & 0.7691& 0.1309& 0.3092& -0.1178& -0.3389& 0.3022& -1.0785 & 2.97\\
    \end{tabular}
\end{table*}

\begin{figure}
\includegraphics[width=\columnwidth]{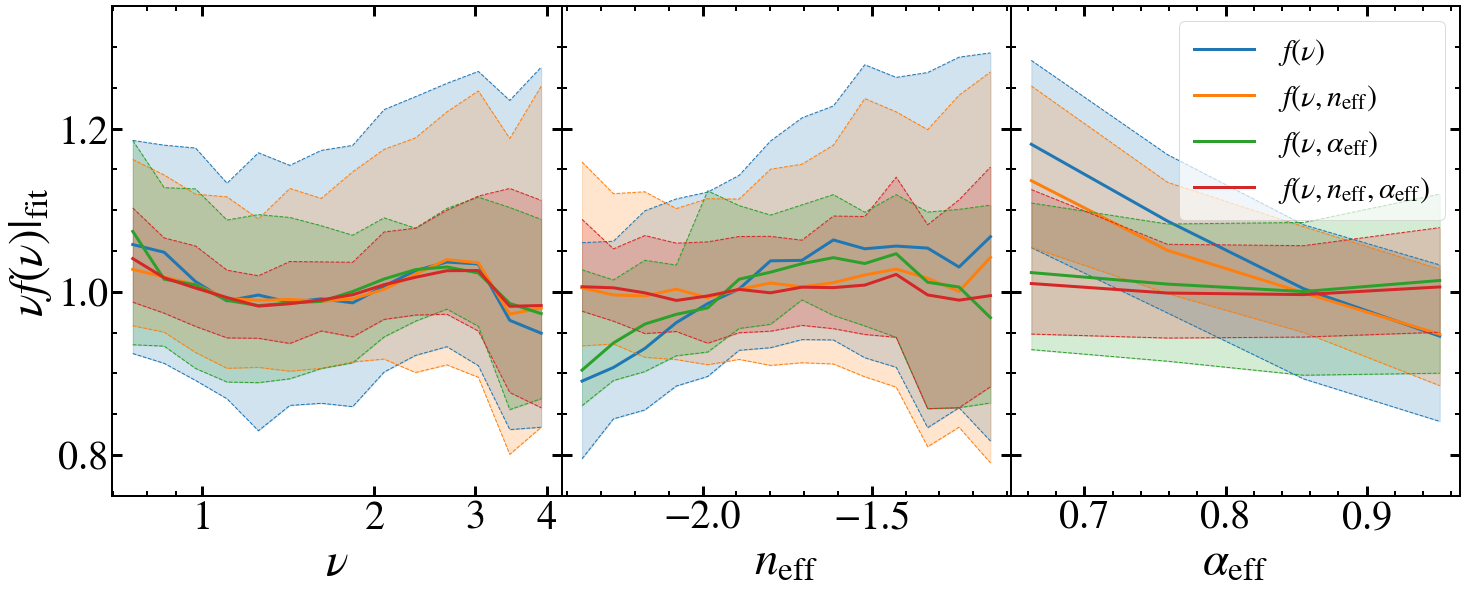}

\hfill
\caption{\label{fig:performance_m200b} Deviations between $\nu\fnu$ measured in our $N$-body simulations and the predictions of the fitting functions developed in this work. Left, middle, and right panels display these deviations as a function of the peak height, $\nu$; the effective power spectrum slope, $\neff$; and the effective growth rate, $\aeff$, respectively. In each panel, shaded regions indicate the region that contains 90\% of our simulated results when employing a fitting function calibrated {\it only} as a function of $\nu$ (blue) or additionally including dependence with respect to $\neff$ (green), $\aeff$ (orange), or both of them (red). }
\end{figure}

We now asses how well this model is able to describe our calibrating data. In Figure~\ref{fig:performance_m200b} we display the ratio of measured $\nu\fnu$, in all our simulations at all three redshifts, to their corresponding predictions of Equation~\ref{eq:x} (red). In the left, middle, and right panels we display the residuals as a function of $\nu$, $\neff$, and $\aeff$, respectively. In all cases, solid and dashed lines display the mean and the median, whereas the shaded areas denote the regions enclosing 90\% of the measurements. Thus, this plot quantifies the overall accuracy of each model in describing the mass function diversity we measured.

We can see that indeed, for the full model, $f(\nu, \aeff, \neff)$, the residuals are smaller than $10\%$ over the whole range of values explored, with no noticeable remaining dependence with either parameter. For comparison, we display also residuals with respect to a version of Equation~\ref{eq:x_nu} where we have measured their parameters to our whole dataset but only adopting dependence with respect to $\nu$ (blue). As expected, in this case, the residuals are significantly larger, reaching variations of $\pm20\%$.

Although the lack of residual dependence with $\aeff$ and $\neff$ is achieved by construction, it is in principle not guaranteed that the amplitude of these residuals decrease significantly. For instance, the mass function could have shown dependence on many more details of the assembly history of haloes and the statistics of peaks than simply on the effective growth rate and power spectrum slope. It is, therefore, remarkable that the residuals in the mass function are all contained within a region of $\pm5\%$, consistent with our statistical uncertainties for the mass function.

In the next section we will explore whether our approach is actually able to describe accurately the mass function in multiple cosmologies currently allowed by observational data.

\section{Validation: Halo abundances as a function of cosmology}\label{sec:validation}

%\lurdes{Both in predicting directly the mass functions and in the scaling correction most of the improvement is coming from $\aeff$, even though $\neff$ provides extra improvement in $n_s$, $\Omega_m$ and $\sigma_8$. Using only $n_s=0.9611$ value for calibrating the fit, the results are a bit poorer but not much. Should we show these things, maybe in an appendix, or it is not very interesting?}

To asses the accuracy of our description for the halo mass function, we will compare its predictions against a suite of simulations with 30 different cosmologies. Each of our simulations evolved $1536^3$ particles inside a box of approximately $L=512\hMpc$. The initial conditions where created using 2nd-order Lagrangian Perturbation theory at $z_{\rm start}=49$ and fixing the amplitude of Fourier modes \citep{AnguloPontzen:2016}. The cosmologies were chosen so that they cover a region of approximately $10\sigma$ around Planck's best fit values. Specifically, they cover the following parameters ranges:

\begin{eqnarray}
\label{eq:par_range}
\sigma_8                  &\in& [0.73, 0.86] \nonumber\\
\Omega_{\rm m}            &\in& [0.23, 0.4] \nonumber\\
\Omega_b                  &\in& [0.04, 0.06] \nonumber\\
n_s                       &\in& [0.92, 0.99]\\
h\,[100\,{\rm km}\,{\rm s^{-1}} {\rm Mpc^{-1}}]  &\in& [0.65, 0.8] \nonumber\\
M_{\nu}\,[{\rm eV}]       &\in& [0.0, 0.4] \nonumber\\
w_{0}                     &\in& [-1.3, -0.7] \nonumber\\
w_{a}                     &\in& [-0.3, 0.3] \nonumber
\end{eqnarray}

Note these simulations not only cover parameters of the minimal $\rm \Lambda CDM$ model, but also neutrino masses, $M_{\nu}$, using the linear response approach of \cite{AliHaimoudBird:2013}; and dynamical dark energy with an equation of state $w(z) = w_0 + (1+z)\,w_a$. The cosmology of each simulation is obtained by changing one cosmological parameter of a fiducial cosmology while keeping the rest fixed. The fiducial cosmology assumes flat geometry, massless neutrinos ($M_{\nu} = 0$), a dark energy equation of state with $w_0 = ˆ'-1$ and $w_a = 0$, an amplitude of matter fluctuations $\sigma_8 = 0.9$, cold dark matter density $\Omega_{\rm cdm}=0.265$, baryon density $\Omega_{\rm b}=0.05$, and normalised Hubble constant $h=0.6$.

\begin{figure}
\centering % \begin{center}/\end{center} takes some additional vertical space
\includegraphics[width=\columnwidth]{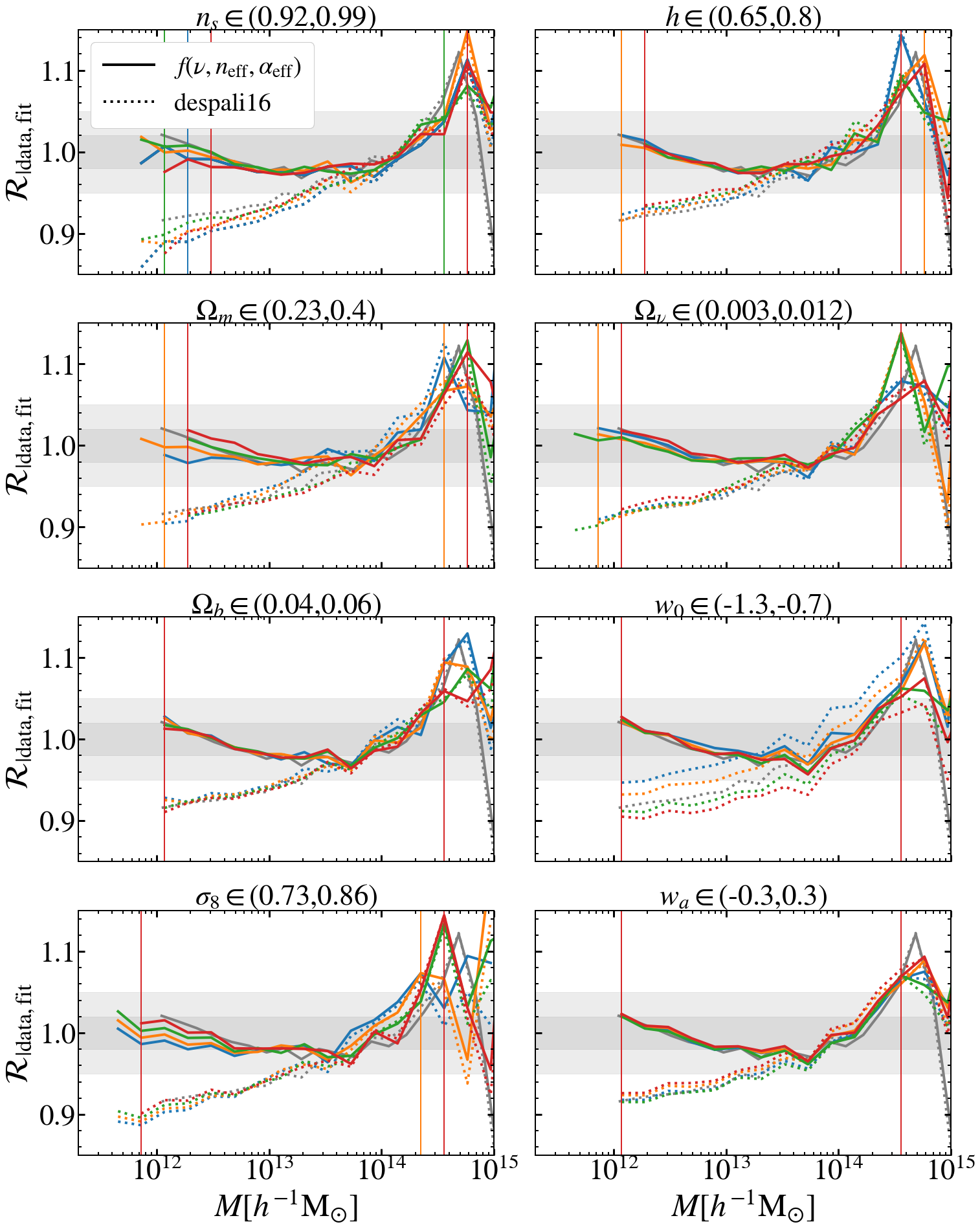}
\hfill
\caption{Comparison between $M_{\rm 200b}$ halo mass functions in multiple cosmologies as measured in $N$-body simulations relative to the fit developed in this work (solid lines) and the model developed in \protect\cite{Despali:2015}(dotted lines), at $z=0$. We display $\mathcal{R}_{|\rm data,\rm fit} = \frac{\rm dn}{\rm dlnM} \left ( \frac{\rm data}{\rm fit} \right)$. The vertical lines display the limits of the bins with at least 400 haloes resolved with more than 200 particles. Each row, from top to bottom, display variations in $n_s$, $\Omega_{\rm m}$, $\Omega_{\rm b}$ and $\sigma_8$ for the first column and  $h$, $\Omega_{\nu}$, $w_0$ and $w_a$ for the second column. The gray lines display the result for the fiducial cosmology. The shaded areas denote regions of $\pm5\%$ and $\pm2\%$.
\label{fig:direct_fit_m200b_z0}}
\end{figure}

\begin{figure}
\centering % \begin{center}/\end{center} takes some additional vertical space
\includegraphics[width=\columnwidth]{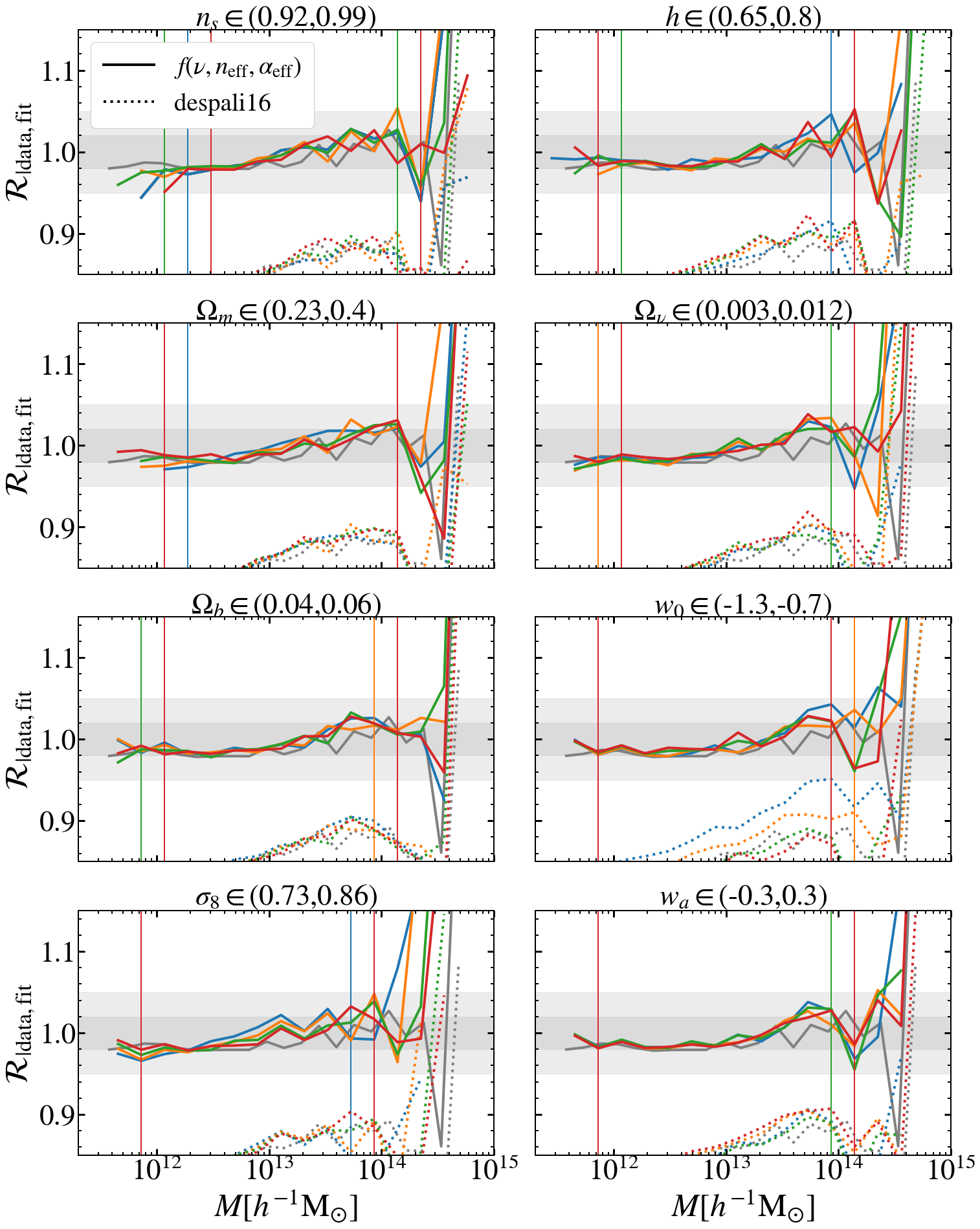}
\hfill
\caption{Same as Figure~\ref{fig:direct_fit_m200b_z0} at $z=1$.\label{fig:direct_fit_m200b_z1}}
\end{figure}

In Figure~\ref{fig:direct_fit_m200b_z0} and ~\ref{fig:direct_fit_m200b_z1} the solid lines represent the measured mass functions of the $30$ simulations relative to the predictions of the main model developed in this work (Eq.~\ref{eq:x}). We show the limits of the bins with at least 400 haloes with resolved with more than 200 particles as vertical lines.

In each row we show the mass functions of the cosmologies where we vary one cosmological parameter keeping the rest fixed. In Figure~\ref{fig:direct_fit_m200b_z0} we display the results at $z=0$,  and in Figure~\ref{fig:direct_fit_m200b_z1} the results at $z=1$. For , we also show the residuals respect the model developed in \cite{Despali:2015} as dotted lines, which assumes universality of the mass function. We recall that the functional dependence on $f(\nu)$ is the same in both models, while in our model we have added extra dependences on $\neff$ and $\aeff$ in order to capture the effect of growth history on the mass function.

At $z=0$, our model describes the low mass end of the mass function at an accuracy of 3\%, while \cite{Despali:2015} predicts that haloes are 10\% more abundant. This may be a consequence of using different group finders, SO and FoF respectively. For haloes with masses above $M>10^{14}\hMsun$, there seems to be an underprediction of our fitting function. To investigate this, we have compared our predictions against the simulations of \cite{Angulo:2020}, which feature the same mass resolution as our test suite but on a volume 27 times larger. Although not shown here, in such case we find an agreement to better than $5\%$ up to $10^{15}\hMsun$. Combined with the good agreement of our predictions with those of \cite{Despali:2015}, we speculate that there is a systematic over prediction of the abundance of haloes in our test sims for $M>10^{14}\hMsun$, which could be caused by finite-volume effects.

At $z=1$, the redshift evolution of the mass function is evident. Even if at $z=0$ \cite{Despali:2015} is a good description to the mass function, at $z=1$ it overpredicts the abundances for more than 10\%. Our model captures this and yields results that are accurate within 5\% at all masses considered.

Compared to the redshift evolution, the cosmology dependence of the mass function seems to be weak. However, the mass functions of $w_0$ cosmologies present strong deviations from universality. The scatter of the ratio with respect to \cite{Despali:2015} is of $\sim 5\%$ among cosmologies with different $w_0$ values at both redshifts. After taking into account the dependences on $\aeff$ and $\neff$, this scatter vanishes. We emphasise that we have only used $\rm \Lambda CDM$ cosmolgies to calibrate the fit, and so $\aeff$ and $\neff$ are physically meaningful proxies of the non-universality of the mass function.

It is interesting to notice that the largest part of the improvement is obtained when adding $\aeff$ to the universal description. This is expected, because as discussed in Section \ref{sec:nonuniversality}, the deviations from universality correlate much stronger with $\aeff$ than with $\neff$.

\section{Application: Improving the accuracy of cosmology-rescaling methods}\label{sec:scaling}

\begin{figure}
\centering % \begin{center}/\end{center} takes some additional vertical space
\includegraphics[width=\columnwidth]{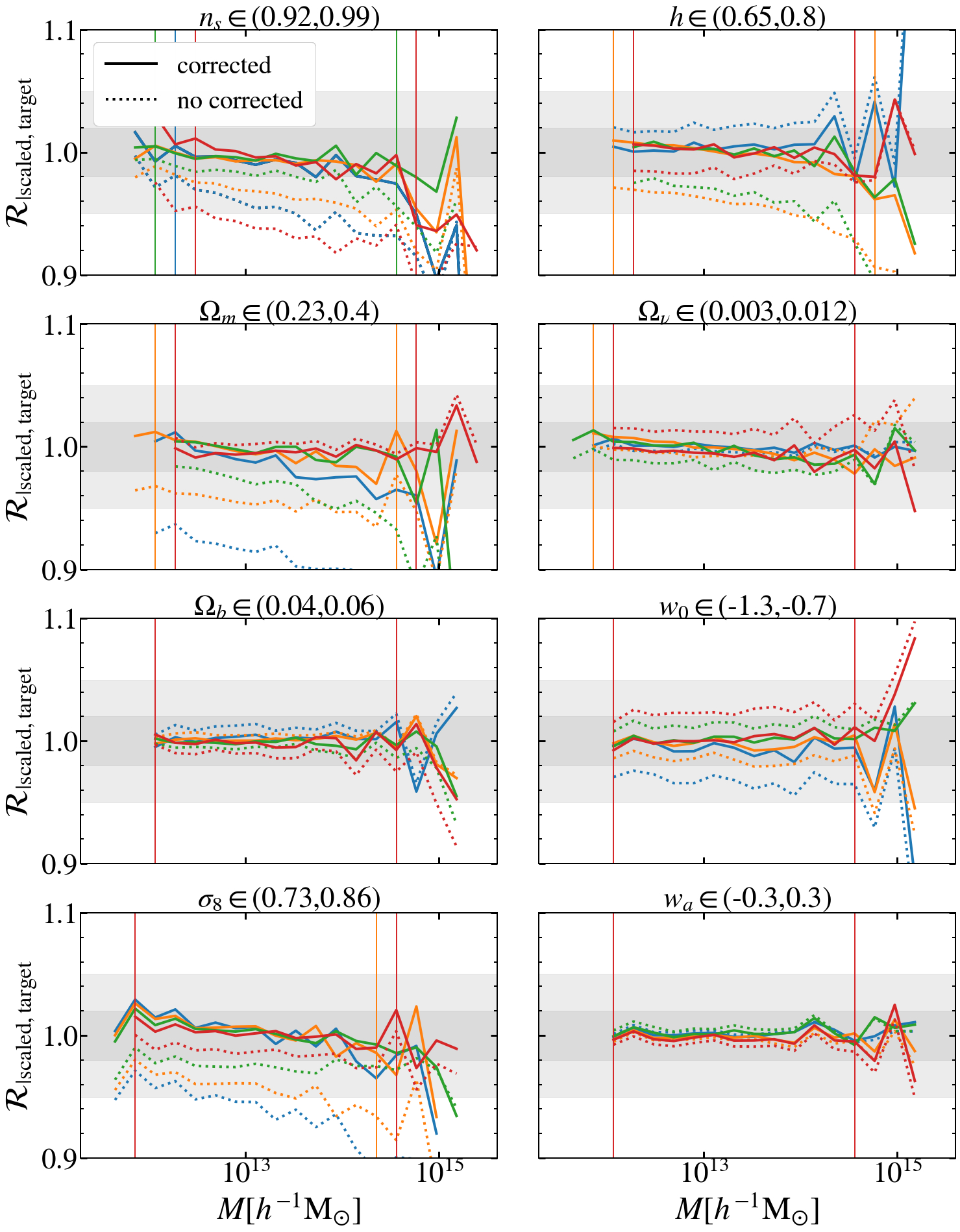}
\hfill
\caption{Comparison between $M_{\rm 200b}$ halo mass functions in multiple cosmologies as measured in $N$-body simulations relative to that in cosmology-rescaled simulations at $z=0$. Specifically, we display $\mathcal{R}_{|\rm scaled,\rm target} = \frac{\rm dn}{\rm dlnM} \left ( \frac{\rm scaled}{\rm target} \right)$ .The vertical lines display the limits of the bins with at least 400 haloes resolved with more than 200 particles. Each row, from top to bottom, display variations in $n_s$, $\Omega_{\rm m}$, $\Omega_{\rm b}$ and $\sigma_8$ for the first column and  $h$, $\Omega_{\nu}$, $w_0$ and $w_a$ for the second column. In each panel we show results before (dotted lines) and after (solid lines) applying our additional correction accounting for dependence on growth history, as indicated by the legend (see text for details). The shaded regions denote regions of $\pm5\%$ and $\pm2\%$.
\label{fig:comp_scaling_m200b}}
\end{figure}

\begin{figure}
\centering % \begin{center}/\end{center} takes some additional vertical space
\includegraphics[width=\columnwidth]{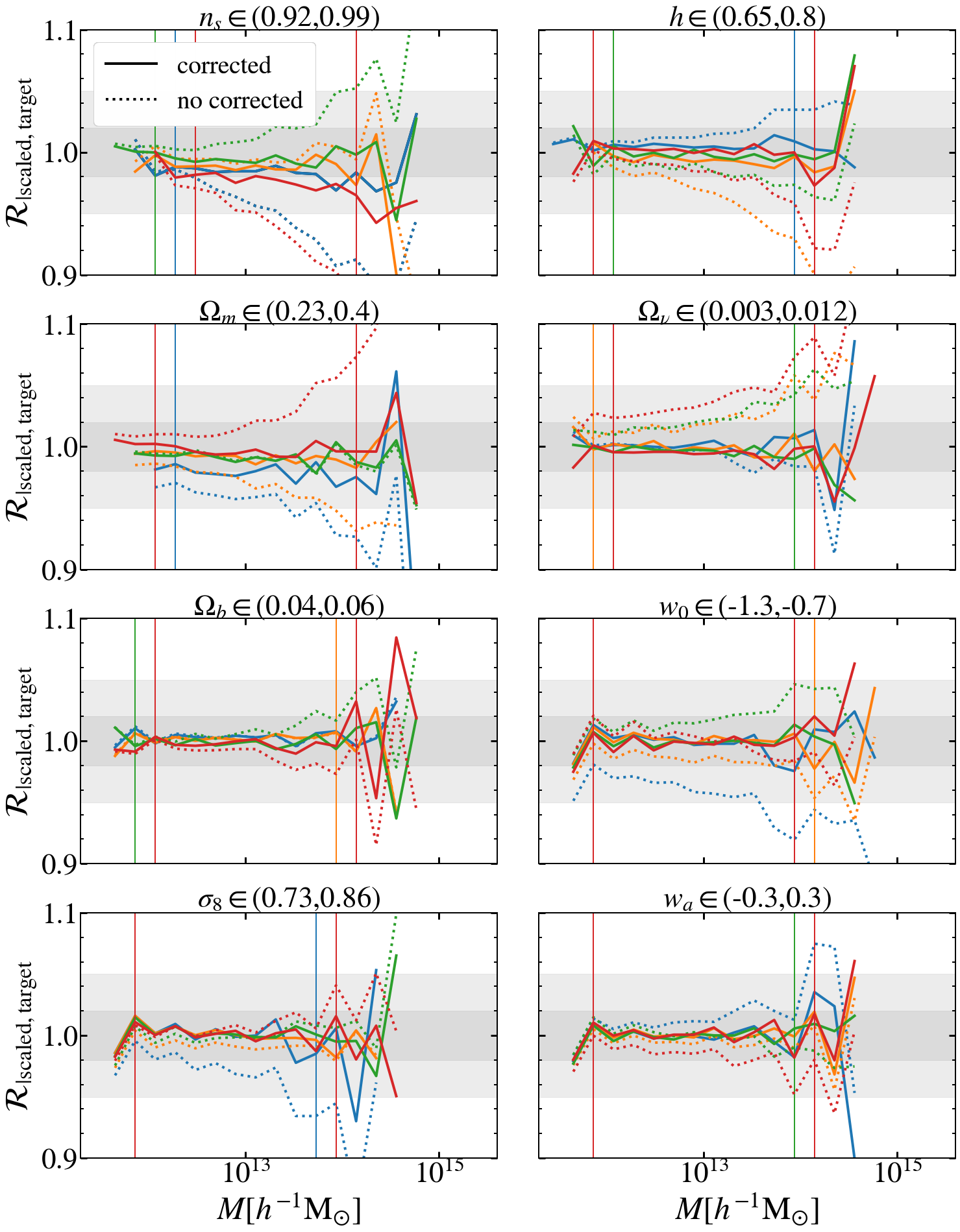}
\hfill
\caption{Same as Fig.\ref{fig:comp_scaling_m200b} but at $z=1$.  \label{fig:comp_scaling_m200b_z1}}
\end{figure}

To approach an optimal exploitation of current and future observations of the abundance of dark matter haloes and the clustering of galaxies, very accurate theoretical predictions for these quantities are required. Although fitting functions and calibrated recipes are indeed extremely valuable, they fall short in providing correlations among different observables or the full three dimensional distribution of clusters of galaxies. One option to obtain those predictions is to employ cosmological $N$-body simulations together with cosmology rescaling algorithms.

The basic idea of cosmology rescaling is to employ a few simulations carried out adopting specific cosmological parameters, and then manipulate their outputs so that represent nonlinear structure in any other set of cosmologies. These algorithms have been extensively discussed and tested in \cite{AnguloWhite:2010,AnguloHilbert:2015,Ruiz:2011,Mead:2014_a,Mead:2014_b,Renneby:2018}. In particular, \cite{Zennaro:2019} showed these are applicable to cases of massive neutrinos, and \cite{Contreras:2020} showed that the clustering of dark matter and dark matter haloes and subhaloes can be obtained to better than 3\% accuracy from large to very small scales ($0.01 < k/\ihMpc < 5$). This technique has been recently employed by \cite{Angulo:2020} to predict the nonlinear power spectrum as a function of cosmology, by \cite{Arico:2019,Arico:2020} to model the effect of baryonic physics and predict the suppression of the power spectrum due to baryons, and by \cite{Zennaro:2021} to model the clustering of biased tracers.

In Figure~\ref{fig:comp_scaling_m200b} we compare the performance of cosmology rescaling in predicting the halo mass functions. Specifically, we compare measurements in the 30 simulations described in the previous section to the halo mass function after rescaling one simulation. We refer the reader to \cite{Contreras:2020} for details on how the cosmologies of these simulations were chosen.

In the original cosmology-rescaling, the simulation volume and the particle mass are rescaled by a single factor, found by minimizing the difference in the linear mass variance in the target and rescaled cosmologies. In this operation the number of particles in each halo is left invariant. Using this recipe the scaling algorithm sets the same linear density field in the rescaled and target simulations, which is equivalent to assuming the universality of the mass function. However, in this work we have shown that the mass function depends not only on the linear density field but also on the entire growth history. In fact, we can see this effect in the dotted lines of Figure~\ref{fig:comp_scaling_m200b}, where we compare the mass functions from the rescaled simulations with the target ones. By assuming universality of the mass function, the rescaled $M_{\rm 200b}$ mass functions differ from the target mass functions up to $10\%$ in some cosmologies at $z=0$.

Our model for the dependence of the halo mass function on growth history gives us the possibility to construct an additional correction for cosmology rescaling by taking into account the different growth histories the target and rescaled cosmologies have gone through. Specifically, we first predict the rescaled and target mass functions. In the target cosmology case, the prediction is straightforward. In the rescaled original cosmology case, we compute the expected mass function of the original cosmology once we have applied the corresponding mass and length scalings, i.e, once we have set the linear density field equal to the target cosmology's linear density field. As we have seen in Section~\ref{subsection:crossmatching}, the growth-history affects the masses of the haloes, not the abundance of them. Thus, the difference between target and rescaled mass functions is given by a change in mass, which we find by mapping the rescaled masses to the target masses where the abundances are the same. Therefore, for a given pair of original-target cosmologies, we can predict a halo-by-halo mass correction that describes the effect of the non-universality of the halo mass function.

We show the results after applying this correction as solid lines in Figure~\ref{fig:comp_scaling_m200b}. We can see that in all cosmologies, the accuracy of the predictions improve in a clear manner. At $z=0$ the differences are in most of the cases smaller than 2\%. Note that our model, calibrated on simulations where we only vary $\Omega_{\rm m}$ and $n_s$, is able to capture the non-universality of general cosmologies, even in beyond $\rm \Lambda CDM$ cosmologies with massive neutrinos and dynamical dark energy included. As seen in Figure~\ref{fig:comp_scaling_m200b_z1}, at $z=1$ the accuracy is as good as for $z=0$, reaching $\pm 1-2\%$ over the full range of masses where we can measure the mass function accurately.

\section{Summary and Conclusions}
\label{sec:conclusions}
In this paper we have studied the non-universality of the halo mass function. We have run simulations with very extreme cosmologies to maximise the deviations from universal behaviour and we have shown that the halo masses are affected by the entire growth history. As a consequence, given the same linear density field in two different cosmologies, the halo mass functions are different.

In order to shed light in the origin of the non-universality of the mass functions, we have crossmatched haloes of different cosmologies that share the same linear density field and we have compared their density profiles. Generally, we have observed that all the density profiles up to very large radii are affected by the growth history  of the haloes. Furthermore, the physical boundaries of haloes selected with density criteria are subject to pseudo evolution, and correspond to different physical radii for different cosmologies. This effect is more pronounced for overdensities defined with respect to the critical density of the universe. Therefore, different mass definitions yield mass functions with different dependences on redshift and cosmology.

We have modelled the non-universality of the mass function adding two additional parameters other than the peak-height $\nu$: the effective growth rate, $\aeff$, and the local slope of the power spectrum, $\neff$. Using a total of 8 free parameters, our model captures the non-universality and can lower the scatter on the halo mass functions in all the cosmologies considered from $\pm 20\%$ to about $\pm 5\%$ up to $z=1$. In the literature, the redshift evolution of the mass function is typically parametrised explicitly within a fiducial cosmology \citep[see e.g.][]{Tinker:2008}. On the contrary, here we have modelled simultaneously the cosmology and redshift non-universality of the mass function by using physically motivated parameters.

We have tested our model on an independent set of simulations of 30 different cosmologies, including massive neutrinos and dynamical dark energy. By considering the $\aeff$ and $\neff$ dependences, we have been able to reproduce the halo mass functions within a $5\%$ accuracy in all the cosmologies up to $M \sim 5 \times 10^{14} \hMsun$ until $z=1$. We emphasise that the simulations that we have used to calibrate the models have been run with $\rm \Lambda CDM$ cosmologies. Thus, it is not a trivial result that our model is able to describe the halo mass functions within, for instance, cosmologies that include massive neutrinos or dynamical dark energy.

As an application of our model, we have applied it together with the cosmology rescaling method presented in \cite{AnguloWhite:2010}. We have found that the accuracy in the scaling of the halo mass function improves from $10\%$ to $2\%$ in all cosmologies including dark energy and massive neutrinos, mostly because of the dependency on the growth rate.

There are many paths that we would like to explore in future works. It is well known that baryonic processes alter in non-trivial way the halo mass function. In particular, astrophysical feedback ejects a large amount of gas outside the haloes boundaries, and therefore haloes result less massive, even by when including a baryonic modelling \citep[see e.g.][]{Castro:2021, Debackere:2020}.
We plan to extend our formalism to include the effect of baryons on the halo mass function, by using the so-called baryonification technique \citep{Schneider:2015, Arico:2019}. By combining it with cosmological-rescaling algorithms, we will construct an emulator of the halo mass functions, as a function of cosmological and astrophysical parameters.

In the near future, more precise and accurate predictions of the halo mass function will be necessary in order to fully exploit the data of the future surveys. As an example, \cite{artis:2021} estimated that, only considering the precision of the parameters of the fitting functions in the analysis (i.e assuming universality of the mass function), an improvement from 30\% to 70\% is required. This framework provides us with a very accurate fit of the halo mass function, which can be eventually exploited to directly compare against observed clusters count, from optical, X-ray or Sunyaev-Zel'dovich surveys. Thus, we anticipate that this model will be of great value value in constraining the cosmological parameters of the Universe.

\section*{Acknowledgments}

LO acknowledges the Summer Internship Program of the Donostia International Physics Center. The authors acknowledge the support of the ERC-StG number 716151 (BACCO). SC acknowledges the support of the ``Juan de la Cierva Formaci\'on'' fellowship (FJCI-2017-33816). The authors acknowledge computing resources at MareNostrum-IV and the technical support provided by Barcelona Supercomputing Center (RES-AECT-2019-2-0012, RES-AECT-2020-3-0014). The authors thank Jens St\"uecker for the visualization routine used in this work.  

%%%%%%%%%%%%%%%%%%%%%%%%%%%%%%%%%%%%%%%%%%%%%%%%%%
\section*{Data Availability}

The data underlying this article will be shared on reasonable request to the corresponding author.



\bibliographystyle{mnras}
\bibliography{bibliography} % if your bibtex file is called example.bib

%\begin{thebibliography}{99}
%\end{thebibliography}

%%%%%%%%%%%%%%%%%%%%%%%%%%%%%%%%%%%%%%%%%%%%%%%%%%

%%%%%%%%%%%%%%%%% APPENDICES %%%%%%%%%%%%%%%%%%%%%

\appendix
\section{Cosmology and redshift dependent critical density}\label{sec:crit_z}
\begin{figure}
\centering % \begin{center}/\end{center} takes some additional vertical space
\includegraphics[width=.8\columnwidth]{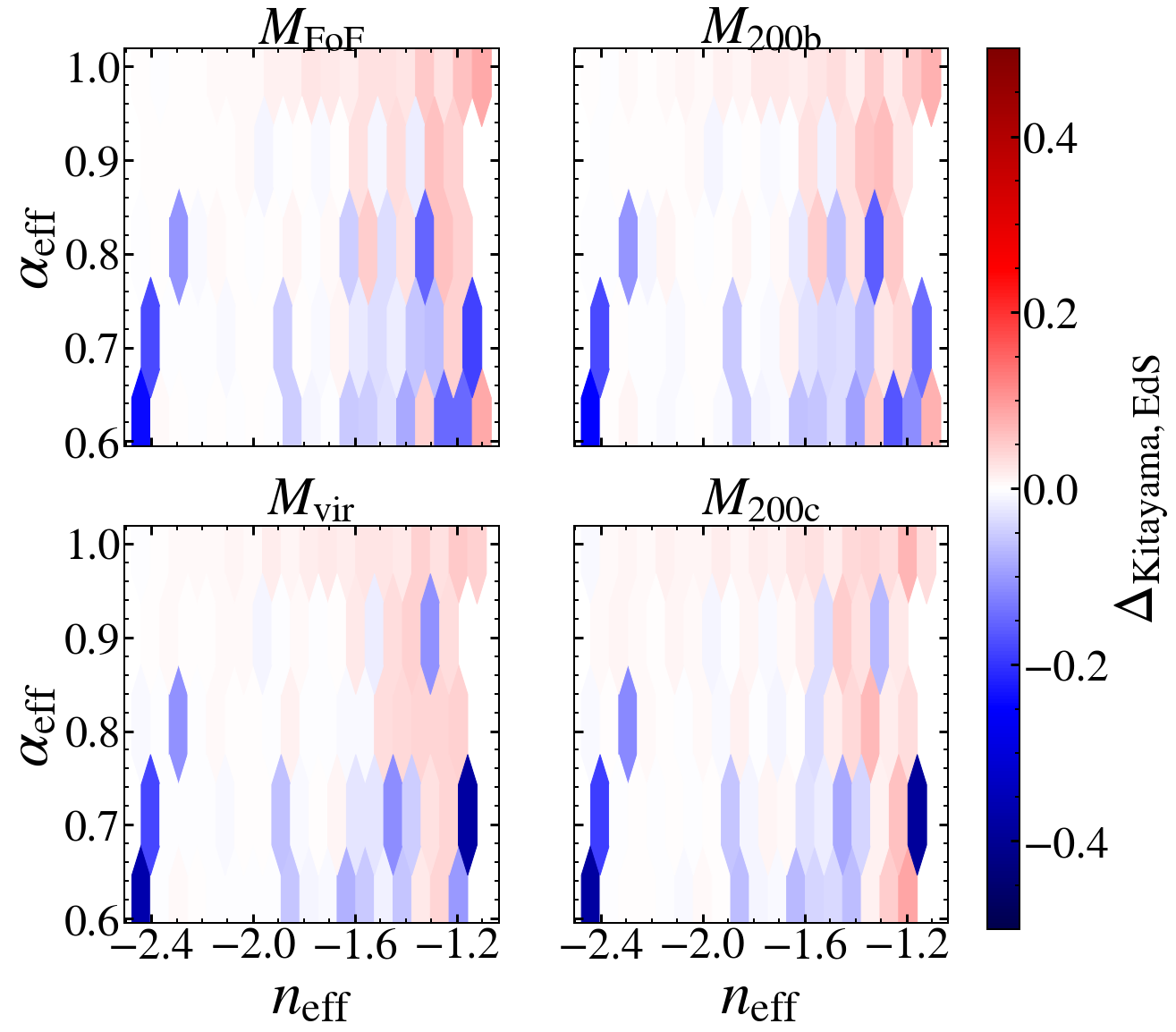}
\hfill
%\includegraphics[width=.45\textwidth]{}
% "\includegraphics" is very powerful; the graphicx package is already loaded
\caption{\label{fig:mass_function_delta_kitayama} The difference of the deviations of the mass function respect to the mean computed in each $\nu$ bin between the mass functions obtained with the critical density for collapse presented in \citep{Kitayama:1996} and with the critical density corresponding to a universe with only matter. $\Delta_{\rm Kitayama, EdS} = (f(\nu)_{\rm Kitayama} - f(\nu)_{\rm EdS}) / f(\nu)_{\rm EdS}$ where $f(\nu) = \nu f(\nu)/< \nu f(\nu)>$.}
\end{figure}

\begin{figure}
\centering % \begin{center}/\end{center} takes some additional vertical space
\includegraphics[width=.8\columnwidth]{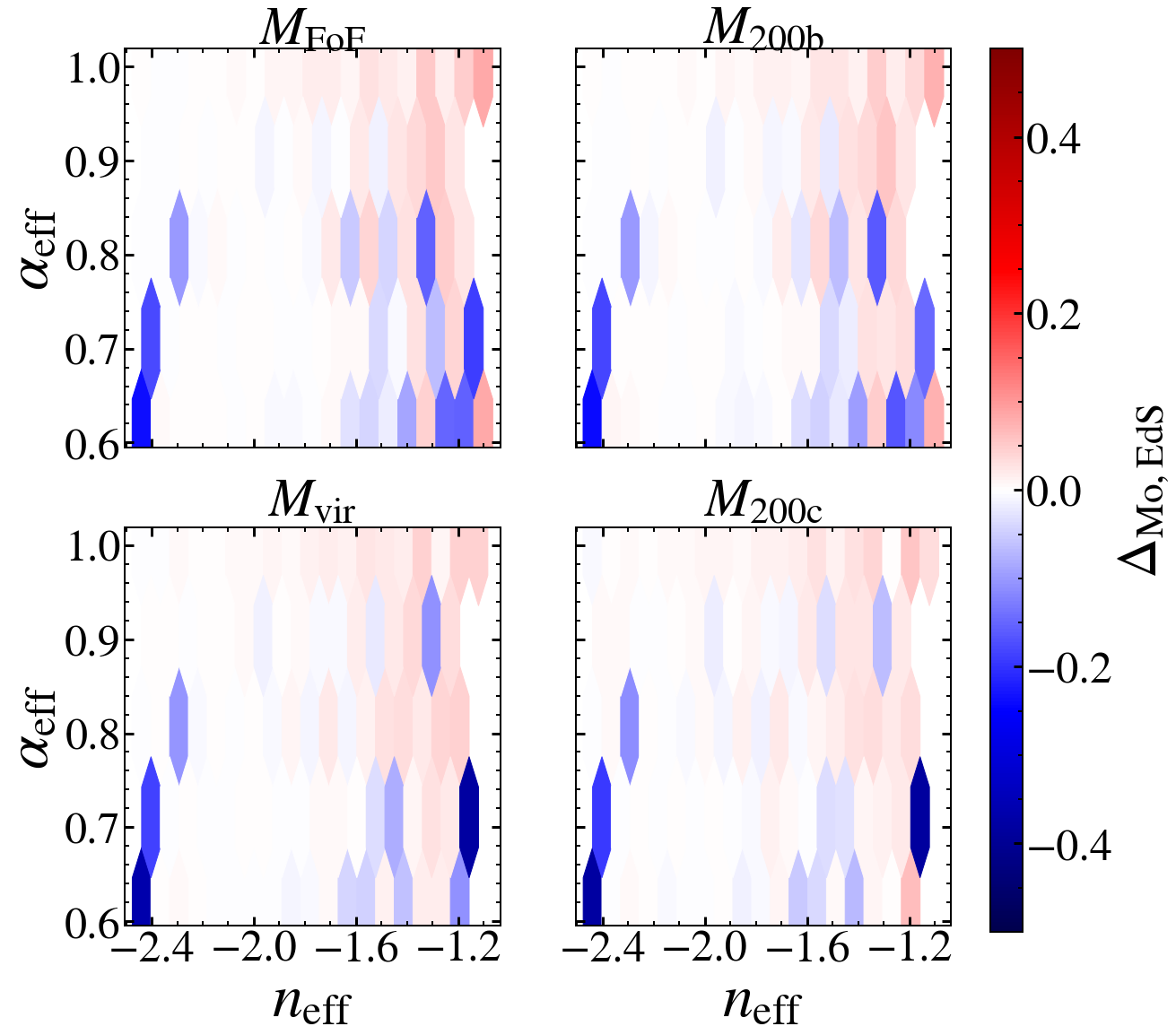}
\hfill
\caption{\label{fig:mass_function_delta_mo} Same as Figure~\ref{fig:mass_function_delta_kitayama} but for the critical density computed following  \citep{Mo:2010}.}
\end{figure}

In this appendix we show the effect of taking into account the redshift dependence of the critical density for collapse on the non-universality of the mass function. Specifically, we compute the relative difference of the deviations in each $\nu$ bin between the mass functions with $\delta_c(z)$ and $\delta_c = 1.686$.

In Figures~\ref{fig:mass_function_delta_kitayama} and~\ref{fig:mass_function_delta_mo} we show these relative differences for two different approaches of computing $\delta_c(z)$. We see that, for most of the cases they do not exceed the $\pm 10\%$, while the deviations around the mean are much stronger, as we can see in Figure~\ref{fig:mass_function_delta}.

\section{Redshift correction}\label{sec:redsh_corr}
\begin{figure}
\begin{tabular}{cc}
  \includegraphics[width=0.47\columnwidth]{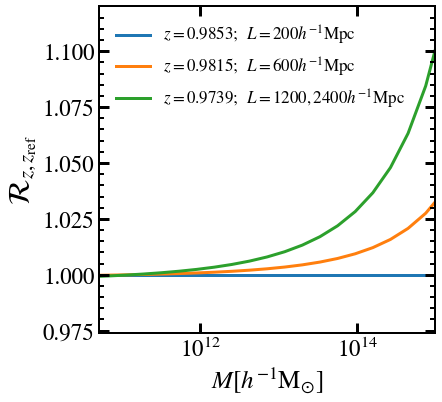} &  \includegraphics[width=0.47\columnwidth]{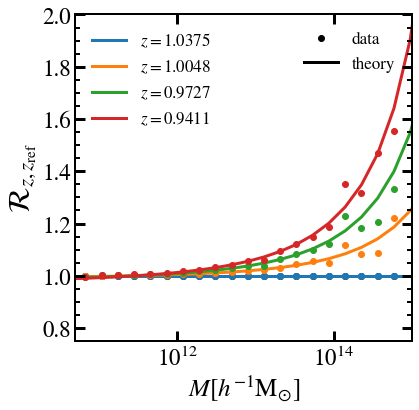} \\
\end{tabular}
\caption{\label{fig:redshift_correction} The ratio of the differential mass function between two expansion factors. \textit{Left panel:} Predicted ratios for the output expansion factors corresponding to different box sizes of our simulation set. The reference redshift is $z_{\rm ref} = 0.9853$. \textit{Right panel:} Predicted and measured ratios for the expansion factors listed in the legend. The reference redshift is $z_{\rm ref}=1.0375$.}
\end{figure}

For different box sizes the redshifts of the snapshots vary slightly. This effect is more pronounced at high redshift, where the difference of the output redshifts of different boxes can reach $\Delta z \sim 0.01$. In this time lapse the mass functions may have evolved, therefore, when combining different box sizes we may be introducing some bias in our data set. The left panel of Figure~\ref{fig:redshift_correction} displays the expected ratio of the differential mass functions between the output redshifts of the different boxes around $z=1$. At $M \sim 10^{15} \hMsun$, the differences can reach 10\%.

In order to test whether the predicted evolution of the mass function is accurate, we make use of a simulation presented in Section~\ref{sec:scaling}, for which we have many snapshots. In the right panel of Figure~\ref{fig:redshift_correction} we display the predicted and measured ratios for the redshifts listed in the legend. We can see that the predictions are a good description of the data. For other cosmologies the results are similar.

Thus, we proceed to correct the mass functions of the big box sizes in the following way,
\begin{equation}
\frac{dn}{d\ln M}|_{\rm corrected} (z_{\rm ref}) =  \frac{dn}{d\ln M}|_{\rm measured} (z) \times \frac{f(M,z_{\rm ref})}{f(M,z)},
\end{equation}
where $f(M,z)$ is some model for the differential mass function.

\section{Extension to other mass definitions}\label{sec:extension_to_masses}

In this appendix we present the main results of our modeling with other mass definitions. These results are analogous to what already described for $M_{\rm 200b}$ mass functions.

In general, as seen in Figure~\ref{fig:mass_function_delta} all mass functions show clear correlations with $\neff$ and $\aeff$ for a given $\nu$.  Therefore, we keep the quadratic and linear functional forms for $\neff$ and $\aeff$ in our model (Eq. ~\ref{eq:x_neff} and \ref{eq:x_aeff}), but slightly change the functional form of the peak-height dependence for the different mass definition.  For $M_{\Delta}$ mass functions we use the functional form used in \cite{Despali:2015} (Eq. ~\ref{eq:x_nu}). However, the functional form used in \cite{Angulo:2012} is more suited to describe $M_{\rm FoF}$ mass functions. Hence, for this mass definition we replace $f_1$ with
\begin{equation}\label{eq:x_nu_tilda}
\tilde{f}_1(\nu) = A\,( b\,\nu^c + 1 )\,\exp(-d\,\nu^2)
\end{equation}
where $\{A, b, c, d\}$ are the free parameters of the $\nu$ dependence of our model. By applying the methodology explained in Section~$\ref{sec:modelling}$, we have obtained the best-fit parameters listed in Tables~\ref{tab:best_fit_pars_fof},~\ref{tab:best_fit_pars_m200c}  and~\ref{tab:best_fit_pars_mvir} for $M_{\rm FoF}$, $M_{\rm 200c}$ and $M_{\rm vir}$ mass functions respectively.

\begin{figure}
\includegraphics[width=\columnwidth]{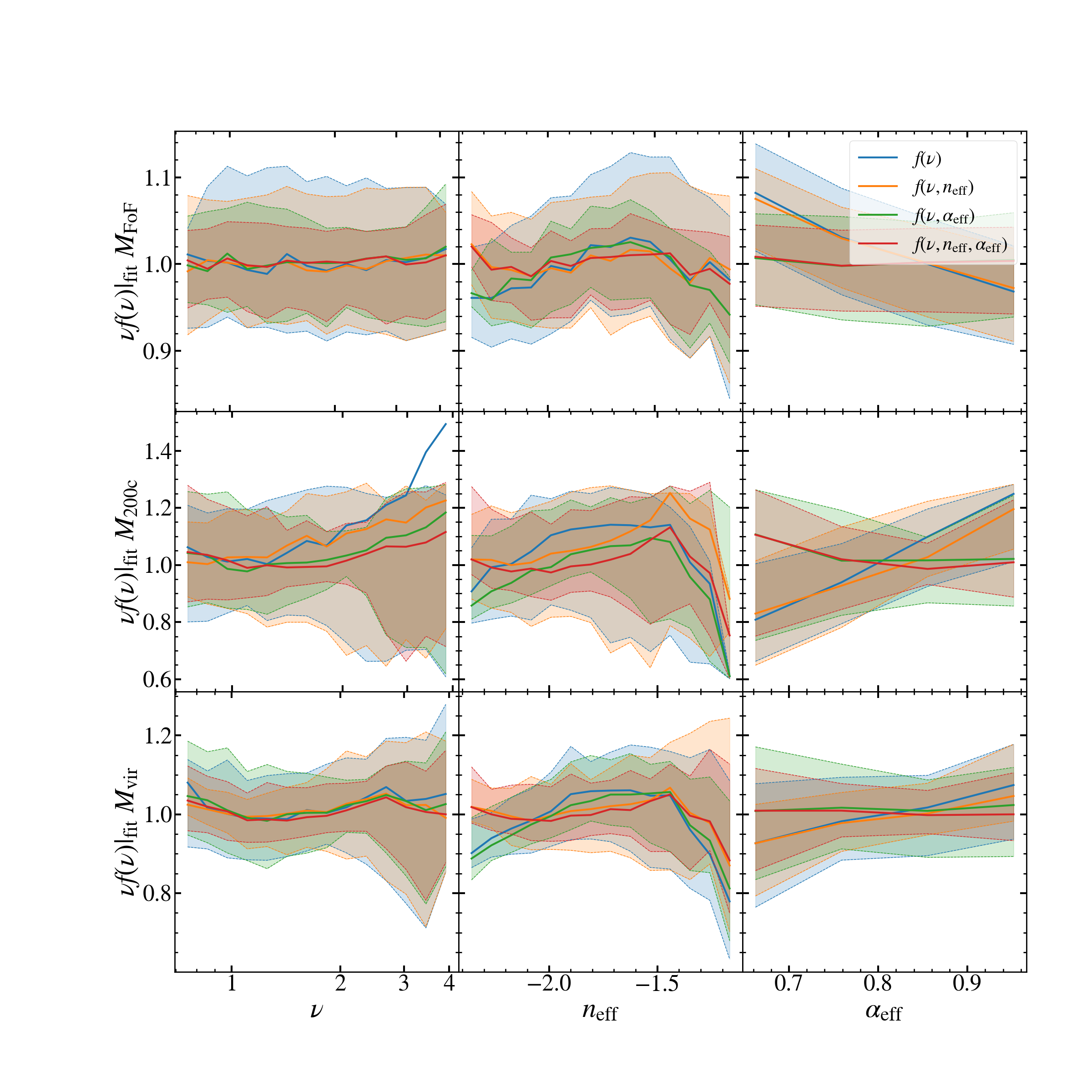}
\caption{\label{fig:fit_all_masses} Ratio of the measured mass functions respect to our model. The ratios are displayed against $\nu$, $\neff$ and $\aeff$ values in the columns. In the rows we show the results for $M_{\rm FoF}$, $M_{\rm 200c}$ and $M_{\rm vir}$ mass functions.}
\end{figure}

In Figure~\ref{fig:fit_all_masses} we show the performance of the model for each mass definition. For $M_{\rm FoF}$ mass functions the residual scatter is consistent with the intrinsic uncertainties of our fit. However, even if the scatter is reduced significantly, for $M_{\rm vir}$ and $M_{\rm 200c}$ the description is not as good as in the other cases. We think that, for both the cases, this may be a consequence of the pseudo-evolution of the boundary definition.

\begin{table*}
  \centering
    \caption{\label{tab:best_fit_pars_fof} A table listing the best fit parameters of our $M_{\rm FoF}$ fitting functions.}
  \begin{tabular}{r r r r r r r r r r}
  	& $A$ & $b$ & $c$ & $d$ & $n_0$ & $n_1$ & $n_2$ & $a_0$ & $a_1$\\
	   \hline
    $f_1(\nu)$ & 0.231 & 1.6871 & 1.7239 & 1.1092& -- & -- & -- & -- & -- \\
    $f_1(\nu)f_2(\neff)$ &0.2297&  1.6824 & 1.6437& 1.0975& -0.1565& -0.4757& 0.6947 & -- & -- \\
    $f_1(\nu)f_3(\aeff)$ &0.2218& 1.8171& 1.6643& 1.1009 & -- &-- &-- &  -0.2908& 1.2186\\
    $f(\nu, \neff, \aeff)$ &0.2276& 1.7692& 1.6249& 1.09& -0.1398& -0.473& 0.3671& -0.3715& 1.6164\
    \end{tabular}
\end{table*}

\begin{table*}
  \centering
  \caption{\label{tab:best_fit_pars_m200c} A table listing the best fit parameters of our $M_{\rm 200c}$ fitting functions.}
    \begin{tabular}{r r r r r r r r r}
  	& $a$ & $p$ & $A_{mp}$ & $n_0$ & $n_1$ & $n_2$ & $a_0$ & $a_1$\\
	   \hline
    $f_1(\nu)$ & 0.833 & 0.1753 & 0.263 & -- & -- & -- & -- & -- \\
    $f_1(\nu)f_2(\neff)$ & 0.769 & 0.2936 & 0.2314 & -0.979 & -3.7864 & -2.4854 & -- & -- \\
    $f_1(\nu)f_3(\aeff)$ & 0.8186 & 0.1796 & 0.2541 & -- &-- &-- & \\
    $f(\nu, \neff, \aeff)$ & 0.7957 & 0.3069 & 0.2549 & -0.502 & -1.79 & -0.8305 & 1.8695 & -0.0937\\
    \end{tabular}
\end{table*}

\begin{table*}
  \centering
  \caption{\label{tab:best_fit_pars_mvir} A table listing the best fit parameters of our $M_{\rm vir}$ fitting functions.}
    \begin{tabular}{r r r r r r r r r}
  	& $a$ & $p$ & $A_{mp}$ & $n_0$ & $n_1$ & $n_2$ & $a_0$ & $a_1$\\
	   \hline
    $f_1(\nu)$ & 0.7814 & 0.0854 & 0.3001 & -- & -- & -- & -- & -- \\
    $f_1(\nu)f_2(\neff)$ & 0.7632 & 0.1867 & 0.2939 & -0.4999 & -1.761 & -0.4741 & -- & -- \\
    $f_1(\nu)f_3(\aeff)$ & 0.7793 & 0.0981 & 0.2951 & -- &-- &-- & 0.3581 & 0.7179\\
    $f(\nu, \neff, \aeff)$ & 0.7693& 0.2074 & 0.2861 & -1.0439& -3.4809& -0.3037& 0.1721& 0.2899\\
    \end{tabular}
\end{table*}

%%%%%%%%%%%%%%%%%%%%%%%%%%%%%%%%%%%%%%%%%%%%%%%%%%

% Don't change these lines
\bsp	% typesetting comment
\label{lastpage}
\end{document}